\documentclass[leqno,twoside]{article}
\usepackage{amsmath,amssymb,amsthm}
\usepackage[margin=2.7cm]{geometry} 
\usepackage{titlesec,hyperref}
\usepackage{fancyhdr}
\usepackage{color}
\fancypagestyle{plain}
{
\fancyhf{}

}

\titleformat{\subsection}{\it}{\thesubsection.\enspace}{1pt}{}

\newtheorem{theo}{Theorem}[section]
\newtheorem{lemm}{Lemma}[section]
\newtheorem{defi}{Definition}[section]
\newtheorem{coro}{Corollary}[section]

\newtheorem{rema}{Remark}[section]
\numberwithin{equation}{section}

\allowdisplaybreaks

\begin{document}

\title{Well-posedness, global existence and blow-up phenomena for an integrable multi-component Camassa-Holm system 
\hspace{-4mm}
}

\author{Zeng Zhang$^1$
\quad Zhaoyang Yin$^2$ \\[10pt]
Department of Mathematics, Sun Yat-sen University,\\
510275, Guangzhou, P. R. China.\\[5pt]
}
\footnotetext[1]{Corresponding author. Email: \it zhangzeng534534@163.com; Tel.:+8613725201644;\\
Address: No. 135, Xingang Xi Road, Guangzhou, 510275, P. R. China}
\footnotetext[2]{Email: \it mcsyzy@mail.sysu.edu.cn}
\date{}
\maketitle
\begin{abstract}
This paper is concerned with a multi-component Camassa-Holm system, which has been proven to be integrable and has peakon solutions. This system includes many one-component and two-component Camassa-Holm type systems as special cases. In this paper, we first establish the local well-posedness and a continuation criterion for the system, then we present several global existence or blow-up results for two important integrable two-component subsystems. Our obtained results cover and improve recent results in \cite{Gui,yan}.\\
\vspace*{5pt}

\vspace*{5pt}
\end{abstract}

\vspace*{10pt}
\tableofcontents
\section{Introduction}
In this paper, we consider the following multi-component system proposed by Xia and Qiao in \cite{xiaqiao}:
\begin{align}\label{s0}
\left\{
\begin{array}{l}
m_{jt}=(m_jH)_x+m_jH+\frac{1}{(N+1)^2}\sum_{i=1}^{N}[m_i(u_j-u_{jx})(v_i+v_{ix})+m_j(u_i-u_{ix})(v_i+v_{ix})],\\[1ex]
n_{jt}=(n_jH)_x-n_jH-\frac{1}{(N+1)^2}\sum_{i=1}^{N}[n_i(u_i-u_{ix})(v_j+v_{jx})+n_j(u_i-u_{ix})(v_i+v_{ix})],\\[1ex]
m_j=u_j-u_{jxx},n_j=v_j-v_{ixx},1\leq j\leq N,\end{array}
\right.
\end{align}
where $H$ is an arbitrary function of $u_j,v_j$, $1\leq j\leq N$, and their derivatives.
The above 2N-component Camassa-Holm system is proved to be integrable in the sense of Lax pair and infinitely many conservation laws in \cite{xiaqiao}, where its peakon solutions for the case $N=2$ are also obtained.

Since $H$ is an arbitrary function of $u_j,v_j$, $1\leq j\leq N$, and their derivatives, thus Eq.(\ref{s0}) is actually a large class of systems. As $N=1$, $v_1=2$ and $H=-u_1$, Eq.(\ref{s0}) is reduced to the standard Camassa-Holm (CH) equation
\begin{align}
  m_t+um_x+2u_xm=0,~~~m=u-u_{xx},
\end{align}
which was derived by Camassa and Holm \cite{Camassa} in 1993 as a model for the unidirectional propagation of shallow water waves over a flat bottom. The CH equation, also as a model for the propagation of axially symmetric waves in hyperelastic rods \cite{Dai}, has a bi-Hamiltonian structure \cite{Constantin,Fokas1} and is completely integrable \cite{Camassa,Constantin1}. One of the remarkable properties of the CH equation is the existence of peakons. One can refer to \cite{Alber,Camassa,Constantin2,Constantin3,Cao} for the existence of peakon solitons and multi-peakons. The Cauchy problem and initial boundary problem of the CH equation has been studied extensively: local well-posedness \cite{Constantins,Constantins4,Danchin,Liy,Rodr¨ªguez-Blanco,E-Y,E-Y1}, global strong solutions \cite{ConstantinA,Constantins,Constantins4,E-Y,E-Y1}, blow-up solutions in finite time  \cite{ConstantinA,Constantins,Constantins5,ConstantinsM,Liu,E-Y,E-Y1} and global weak solutions \cite{Bressan,Constantins6,ConstantinM2,xin}.

As $N=1$ and $H=-\frac{1}{2}(u_1-u_{1x})(v_1+v_{1x})$, Eq.(\ref{s0}) is reduced to the following system proposed by Song, Qu and Qiao in \cite{song}:
\begin{align}\label{sqqq}
\left\{
\begin{array}{l}
m_{t}+\frac{1}{2}\big((u-u_{x})(v+v_{x}) m\big)_x=0,\\[1ex]
n_{t}+\frac{1}{2}\big((u-u_{x})(v+v_{x}) n\big)_x=0.
\end{array}
\right.
\end{align}
The above system is proved to be integrable not only in the sense of Lax-pair but also in the sense of geometry, namely, it describes pseudospherical surfaces \cite{song}. Besides, exact solutions to this system such as cuspons and W/M-shape solitons are also obtained in \cite{song}.

As $N=1$ and $H=-\frac{1}{2}(u_1v_1-u_{1x}v_{1x})$, Eq.(\ref{s0}) is reduced to the following system proposed by Xia and Qiao in \cite{Xia,Qiao2}:
\begin{align}\label{qqqx}
\left\{
\begin{array}{l}
m_{t}+\frac{1}{2}\big((uv-u_{x}v_{x}) m\big)_x-\frac{1}{2}(uv_{x}-vu_{x})m=0,\\[1ex]
n_{t}+\frac{1}{2}\big((uv-u_{x}v_{x}) n\big)_x+\frac{1}{2}(uv_{x}-vu_{x})n=0,
\end{array}
\right.
\end{align}
which describes a nontrivial one-parameter family of pseudo-spherical surfaces. In \cite{Xia,Qiao2}, the authors showed this system is integrable with Lax pair, bi-Hamiltonian structure, and infinitely many conservation laws. They also studied the peaked soliton and multi-peakon solutions to the system. Recently, Yan, Qiao and Yin \cite{yan} studied the local well-posedness for the Cauchy problem of the system and derived a precise blow-up scenario and a blow-up result for the strong solutions to the system.

As $v=2u$, both Eq.(\ref{sqqq}) and Eq.(\ref{qqqx}) are reduced to the following cubic Camassa-Holm equation\begin{align}\label{qio}
  m_t+\big(m(u^2-u_x^2)\big)_x=0,~~~m=u-u_{xx},
\end{align}
which was proposed independently by Fokas \cite{Fokas}, Fuchssteiner \cite{Fuchssteiner}, Olver and
Rosenau \cite{Olevr}, and Qiao \cite{Qiao} as an integrable peakon equations with cubic nonlinearity. Its Lax pair, peakon and soliton solutions, local well-posedness and blow-up phenomena have been studied in \cite{Qiao,ggl2,Gui}.

The aim of this paper is to establish the local well-posedness and a continuation criterion for the Cauchy problem of Eq.(\ref{s0}) in Besov spaces, and present several global existence or blow-up results for the two component subsystems: Eq.(\ref{sqqq}) and Eq.(\ref{qqqx}). Our obtained results cover and improve recent results in \cite{Gui,yan}. Compared with the Camassa-Holm equation, one of the remarkable features of Eq.(\ref{s0}) is that it has higher-order nonlinearities. Thus, we have to estimate elaborately these higher-order nonlinear terms for the study of the local well-posedness and the continuation criterion of Eq.(\ref{s0}) in Besov spaces. Besides, we derive that $\|m(t)\|_{L^1}$ ($\|n(t)\|_{L^1}$) and $\int_{\mathbb{R}}(mv_x)(t,x)dx=
 \int_{\mathbb{R}}(nu_x)(t,x)dx$ are conservation laws for Eq.(\ref{sqqq}) and Eq.(\ref{qqqx}), respectively. The above conservation laws, which have not been derived or used in the associated previous papers \cite{Gui,yan}, are useful and crucial in some blow-up results stated in the following fourth section.

The rest of our paper is then organized as follows. In Section 2, we recall the Littlewood-Paley decomposition and some basic properties of the Besov spaces. In Section 3, we establish the local well-posedness and provide a continuation criterion for Eq.(\ref{s0}).
The last section is devoted to establishing several global existence or blow-up results for Eq.(\ref{sqqq}) and Eq.(\ref{qqqx}).

From now on we always assume that $H=H(u_1,\cdots,u_N,v_1,\cdots,v_N,u_{1x},\cdots,u_{1x},v_{1x},\cdots,v_{1x})$ is a polynomial of degree $l$, $C>0$ stands for a generic constant, $A\lesssim B$ denotes the relation $A \leq CB.$ Since all function spaces in this paper are over $\mathbb{R}$, for simplicity, we drop $\mathbb{R}$ in the notations of function spaces if there is no ambiguity.
\section{Preliminaries}
To begin with, we introduce the Littlewood-Paley decomposition.
 \begin{lemm}\cite{keben}
Let $\mathcal{C}=\{\xi\in{\mathbb{R}}, ~\frac{3}{4}\leq|\xi|\leq\frac{8}{3} \}$ be an annulus. There exist radial functions $\chi$ and $\varphi$ valued in the interval $[0,1]$, belonging respectively to $\mathcal{D}(B(0,\frac{4}{3}))$ and $\mathcal{D}(\mathcal{C})$,  such that
\begin{align*}
\forall \xi \in{\mathbb{R}},~~\chi(\xi)+\sum_{j\geq 0}\varphi(2^{-j}\xi)=1.\end{align*}\end{lemm}
The nonhomogeneous dyadic blocks $\triangle_j$ and the nonhomogeneous low-frequency cut-off operator $S_j$ are then defined as follows:
\begin{align*}&\triangle_ju=0~~if~j\leq-2,~~~~~~~~~~~~~\triangle_{-1}u=\chi(D)u,\\
&\triangle_ju=\varphi(2^{-j}D)u~~if~j\geq0,~~~~S_ju=\sum_{j'\leq j-1}\triangle_{j'}u~~for~j\in \mathbb{Z}.\end{align*}
\begin{defi}\cite{keben}\label{dingyi}
Let $s\in \mathbb{R}$ and $(p,r)\in[1,\infty]^2$. The nonhomogeneous Besov space $B^s_{p,r}$
consists of all $u\in\mathcal{S}'(\mathbb{R})$ such that
$$\|u\|_{B^s_{p,r}}\overset{def}{=}\Big\|(2^{js}\|\triangle_ju\|_{L^p})_{j\in \mathbb{Z}}\Big\|_{l^r(\mathbb{Z})}
<\infty.$$
\end{defi}
Let us give some classical properties of the Besov spaces.
\begin{lemm}\label{Fatou}\cite{keben}
The set $B^s_{p,r}$ is a Banach space, and satisfies the Fatou property, namely, if $(u_n)_{n\in N}$ is a
bounded sequence of $B^s_{p,r}$, then an element $u$ of $B^s_{p,r}$ and a subsequence $u_{\psi(n)}$ exist such that
$$\underset{n\rightarrow\infty}{\lim}~u_{\psi(n)}=u~~in~~\mathcal{S}'~~~and ~~~\|u\|_{B^s_{p,r}}\leq C \underset{n\rightarrow\infty}{\liminf} \|u_{\psi(n)}\|_{B^s_{p,r}}.$$\end{lemm}
\begin{lemm}\label{fm}\cite{keben}
  Let $m\in \mathbb{R}$ and $f$ be an $S^m$-multiplier (i.e. $f: \mathbb{R}\rightarrow \mathbb{R}$ is smooth and satisfies that for each multi-index $\alpha$, there exists a constant $C_{\alpha}$ such that $|\partial^{\alpha}f(\xi)|\leq C_{\alpha}(1+|\xi|)^{m-|\alpha|}, \forall \xi\in \mathbb{R}$). Then the operator $F(D)$ is continuous from $B^s_{p,r}$ to $B^{s-m}_{p,r}$.
\end{lemm}
\begin{lemm}\label{chengji}\cite{Danchin}
(i) For $s>0$ and $1\leq p,r\leq \infty$, there exists $C=C(d,s)$ such that
\begin{align}
  \|uv\|_{B^s_{p,r}}\leq C(\|u\|_{L^\infty}\|v\|_{B^s_{p,r}}+\|v\|_{L^\infty}\|u\|_{B^s_{p,r}}).
\end{align}
(ii) If $1\leq p,r\leq \infty$, $s_1\leq\frac{1}{p}$, $s_2>\frac{1}{p}$ ($s_2\geq\frac{1}{p},$ if $r=1$) and $s_1+s_2>\max\{0, \frac{2}{p}-1\},$ there exists $C=C(s_1,s_2,p,r)$ such that
\begin{align}
  \|uv\|_{B^{s_1}_{p,r}}\leq C\|u\|_{B^{s_1}_{p,r}}\|v\|_{B^{s_2}_{p,r}}.
\end{align}
\end{lemm}
\begin{lemm}\label{gai}\cite{keben,Danchin}
  Let $1\leq p\leq p_1\leq \infty,$ $1\leq r\leq \infty,$ $s>-\min\{\frac{1}{p_1},1-\frac{1}{p}\}.$ Assume $f_0\in B^s_{p,r}, F\in L^1(0,T;B^s_{p,r})$, $v\in L^\rho(0,T;B^{-M}_{\infty,\infty})$ for some $\rho>1$ and $M>0,$ and
  \begin{align*}
  &\partial_xv\in L^1(0,T;B^{\frac{1}{p_1}}_{p_1,\infty}\cap L^\infty),&\textit{if}&~ s<1+\frac{1}{p_1}, \\
  &\partial_xv\in L^1(0,T;B^{s-1}_{p_1,r}),&\textit{if}&~ s>1+\frac{1}{p_1},~\textit{or}~s=1+\frac{1}{p_1}~\textit{and}~r=1.
  \end{align*}
Then the following transport equation
\begin{align}
\left\{
\begin{array}{l}
\partial_tf+v\cdot\nabla f=F\\
f_{|t=0}=f_0,
\end{array}
\right.
\end{align}
has a unique solution $f\in C([0,T];B^s_{p,r})$, if $r<\infty,$ or $f \in L^\infty(0,T;B^s_{p,r})\cap\Big(\bigcap_{s'<s}C([0,T];B^{s'}_{p,r})\Big)$, if $r=\infty.$\\
Moreover, the following inequality holds true:
\begin{align}\label{,,}
\|f(t)\|_{B^s_{p,r}}\leq \|f_0\|_{B^s_{p,r}}+\int_0^t\|F(\tau)\|_{B^s_{p,r}}d\tau+C\int_0^t V'_{p_1}(\tau)\|f(\tau)\|_{B^s_{p,r}}d\tau
\end{align}
with
\begin{align}
V'_{p_1}(t)=\left\{\begin{array}{l}
\|\partial_x v(t)\|_{B^{\frac{1}{p_1}}_{p_1,\infty}\cap L^{\infty}},~if~s<1+\frac{1}{p_1},\\
\|\partial_x v(t)\|_{B^{s-1}_{p_1,r}},~if~s>1+\frac{1}{p_1}~or~s=1+\frac{1}{p_1},~r=1.
\end{array}\right.
\end{align}
\end{lemm}
\section{Local well-posedness}
In this section, we study the local well-posedness for Eq.(\ref{s0}).\\
 To begin with, noticing $(1-\partial_x^2)^{-1}=\frac{1}{2}e^{-|x|}\ast,$ we have the following inequalities which will be frequently used in the sequel:
 \begin{align*}
   &\|u\|_{B^{s}_{p,r}}=\|(1-\partial_x^2)^{-1}m\|_{B^{s}_{p,r}}\thickapprox
   \|m\|_{B^{s+2}_{p,r}},~~~\forall~s\in\mathbb{R},~1\leq p,r\leq \infty.\\
   &\|u\|_{L^\infty}=\|\frac{1}{2}e^{-|x|}\ast m\|_{L^\infty}\leq \|m\|_{L^\infty},\\
 &\|u_x\|_{L^\infty}=\|\frac{1}{2}(-sign(x)e^{-|x|})\ast m\|_{L^\infty}\leq \|m\|_{L^\infty},\\
 &\|u_{xx}\|_{L^\infty}=\|u-m\|_{L^\infty}\leq 2\|m\|_{L^\infty},
 \end{align*}
 where $m=u-u_{xx}.$

 We now rewrite Eq.(\ref{s0}) as follows:
\begin{align}\label{new}
\left\{
\begin{array}{l}
M_t=H(U,U_x) M_x+A(H,H_x)M+B(U,U_x)M,\\
M_{|t=0}=M_0,
\end{array}
\right.
\end{align}
where $M=(m_1,\cdots,m_N,n_1,\cdots,n_N)^T,$ $M_0=(m_{10},\cdots,m_{N0},n_{10},\cdots,n_{N0})^T,$ $U=(u_1,\cdots,u_N,v_1,\cdots,v_N)^T$, $H=H(U,U_x)$ is a polynomial of degree $l$, and
\begin{align*}
A(H,H_x)=\left(
\begin{array}{lllcr}
H_xI_{N\times N}+HI_{N\times N}& 0 \\
 0& H_xI_{N\times N}-HI_{N\times N}
\end{array}
\right),~~
B(U,U_x)=\left(\begin{array}{llcr}
B_{11} & 0\\
0 & B_{22}
\end{array}
\right)
\end{align*}with
\begin{align*}
B_{11}=\frac{1}{(N+1)^2}\left(
\begin{array}{llcr}
(u_1-u_{1x})(v_1+v_{1x}) & \cdots & (u_1-u_{1x})(v_N+v_{Nx})\\
\vdots &\vdots & \vdots\\
(u_N-u_{Nx})(v_1+v_{1x})& \cdots& (u_N-u_{Nx})(v_N+v_{Nx})
\end{array}
\right)
+\sum_{i=1}^{N}[(u_i-u_{ix})(v_i+v_{ix})]I_{N\times N},
\end{align*}
and \begin{align*}
B_{11}=-\frac{1}{(N+1)^2}\left(
\begin{array}{llcr}
(u_1-u_{1x})(v_1+v_{1x}) & \cdots & (u_N-u_{Nx})(v_1+v_{1x})\\
\vdots &\vdots & \vdots\\
(u_1-u_{1x})(v_N+v_{Nx})& \cdots& (u_N-u_{Nx})(v_N+v_{Nx})
\end{array}
\right)
-\sum_{i=1}^{N}[(u_i-u_{ix})(v_i+v_{ix})]I_{N\times N}.
\end{align*}
\subsection{Local existence and uniqueness}
\begin{theo}\label{pose}
Let $1\leq p,r\leq \infty$, $s>\max\{1-\frac{1}{p},\frac{1}{p}\}$, and $M_0\in  B^s_{p,r}.$ Then exists a time $T>0$ such that Eq.(\ref{new}) has a unique solution $M\in L^\infty(0,T;B^s_{p,r})\cap E^{s}_{p,r}(T)$ with
\begin{align*}
  E^{s}_{p,r}(T)\triangleq
  \left\{\begin{array}{l}
  C([0,T];B^s_{p,r})\cap C^1([0,T];B^{s-1}_{p,r}),~\textit{if}~r<\infty,\\
 \bigcap_{s'<s} \Big(C([0,T];B^{s'}_{p,r})\cap C^1([0,T];B^{s'-1}_{p,r})\Big),~\textit{if}~r=\infty.
 \end{array}\right.
\end{align*}
\end{theo}
The proof relies heavily on the following lemma.
\begin{lemm}\label{weiyi}
  Let $1\leq p,r\leq \infty$ and $s>\max\{1-\frac{1}{p},\frac{1}{p}\}$. Suppose that $M^1$ and $M^2$ are two solutions of the Eq.(\ref{new}) with the initial data $M^1_0,M^2_0\in L^\infty(0,T;B^s_{p,r})\cap C([0,T];\mathcal{S}')$. Let $M^{12}=M^1-M^2$, $U^{12}=U^1-U^2$, and $q=\max\{l,2\}$(where $l$ is the polynomial order of $H$). Then, for all $t\in[0,T],$ we have \\
  (1) if $s>\max\{1-\frac{1}{p},\frac{1}{p}\}$, but $s\neq 2+\frac{1}{p}$, then \begin{align}\label{12}
    \|M^{12}(t)\|_{B^{s-1}_{p,r}}\leq \|M^{12}_0\|_{B^{s-1}_{p,r}}e^{C\int_0^t(\|M^1(\tau)\|_{B^s_{p,r}}^q
    +\|M^2(\tau)\|_{B^s_{p,r}}^q+1)d\tau};
  \end{align}
(2) if $s=2+\frac{1}{p}$, then
\begin{align}
    \|M^{12}(t)\|_{B^{s-1}_{p,r}}\leq \|M^{12}_0\|_{B^{s-1}_{p,r}}^\theta (\|M^{1}(t)\|_{B^s_{p,r}}+\|M^{2}(t)\|_{B^s_{p,r}})^{1-\theta}e^{\theta C\int_0^t(\|M^1(\tau)\|_{B^s_{p,r}}^q
    +\|M^2(\tau)\|_{B^s_{p,r}}^q+1)d\tau},
    \end{align}
    where $\theta\in(0,1).$
\end{lemm}
\noindent{Proof}. Let $H^i=H(U^i,U^i_x),$ $A^i=A(H^i,H^i_x),$ $B^i=B(U^i,U^i_x),$ $i=1,2,$ and $H^{12}=H^1-H^2,$ $A^{12}=A^1-A^2,$ $B^{12}=B^1-B^2.$ It is obvious that $M^{12}$ solves the following transport equation
\begin{align*}
  M^{12}_t-H^1 M^{12}_x=F_1+F_2+F_3
\end{align*}
where $F_1=H^{12}M^2_x$ $F_2=(A^1+B^1)M^{12}$ and $F_3=(A^{12}+B^{12})M^2.$\\
We claim that for all $s>\max\{1-\frac{1}{p},\frac{1}{p}\},$ we have
\begin{align}\label{cj}
\|uv\|_{B^{s-1}_{p,r}}\lesssim\|u\|_{B^{s-1}_{p,r}}\|v\|_{B^{s}_{p,r}}.
\end{align}
Indeed, if $s>1+\frac{1}{p},$ then $B^{s-1}_{p,r}$ is an algebra. Thus we have
\begin{align*}
\|uv\|_{B^{s-1}_{p,r}}\lesssim\|u\|_{B^{s-1}_{p,r}}\|v\|_{B^{s-1}_{p,r}}\lesssim\|u\|_{B^{s-1}_{p,r}}\|v\|_{B^{s}_{p,r}}.
\end{align*}
On the other hand, if $\max\{1-\frac{1}{p},\frac{1}{p}\}< s\leq1+\frac{1}{p},$ then applying Lemma \ref{chengji} (ii) with $s_1=s-1$ and $s_2=s$ yields (\ref{cj}).\\
Therefore, for all $s>\max\{1-\frac{1}{p},\frac{1}{p}\},$ noticing the fact that $B^s_{p,r}$ is an algebra, one may infer the following inequalities:
\begin{align*}
\|F_1\|_{B^{s-1}_{p,r}}\lesssim&\|H^{12}\|_{B^{s}_{p,r}}\|M^2_x\|_{B^{s-1}_{p,r}}\\
\lesssim &(\|U^{12}\|_{B^{s}_{p,r}}+\|U_x^{12}\|_{B^{s}_{p,r}})(\|U^1\|_{B^{s}_{p,r}}^{l-1}+
\|U_x^1\|_{B^{s}_{p,r}}^{l-1}+\|U^2\|_{B^{s}_{p,r}}^{l-1}+
\|U_x^2\|_{B^{s}_{p,r}}^{l-1}
)\|M^2_x\|_{B^{s-1}_{p,r}}\\
\lesssim&\|M^{12}\|_{B^{s-1}_{p,r}}(\|M^1\|_{B^{s}_{p,r}}^{q}+\|M^2\|_{B^{s}_{p,r}}^{q}+1),\\
\|F_2\|_{B^{s-1}_{p,r}}\lesssim&(\|A^1\|_{B^{s}_{p,r}}+\|B^1\|_{B^{s}_{p,r}})\|M^{12}\|_{B^{s-1}_{p,r}}\\
\lesssim &(\|U^1\|_{B^{s}_{p,r}}^{l}+
\|U_x^1\|_{B^{s}_{p,r}}^{l}+\|U^1_{xx}\|_{B^{s}_{p,r}}^{l}+\|U^1\|_{B^{s}_{p,r}}^{2}+\|U^1\|_{B^{s}_{p,r}}^{2}
+1)\|M^{12}\|_{B^{s-1}_{p,r}}\\
\lesssim&\|M^{12}\|_{B^{s-1}_{p,r}}(\|M^1\|_{B^{s}_{p,r}}^{q}+\|M^2\|_{B^{s}_{p,r}}^{q}+1),\\
\|F_3\|_{B^{s-1}_{p,r}}\lesssim&(\|A^{12}\|_{B^{s-1}_{p,r}}+\|B^{12}\|_{B^{s-1}_{p,r}})\|M^{2}\|_{B^{s}_{p,r}}\\
\lesssim &(\|U^{12}\|_{B^{s-1}_{p,r}}+
\|U_x^{12}\|_{B^{s-1}_{p,r}}+\|U^{12}_{xx}\|_{B^{s-1}_{p,r}})
(\|U^1\|_{B^{s}_{p,r}}^{l-1}+\|U^1_x\|_{B^{s}_{p,r}}^{l-1}+\|U^1_{xx}\|_{B^{s}_{p,r}}^{l-1}+\|U^2\|_{B^{s}_{p,r}}^{l-1}\\
&+\|U^2_x\|_{B^{s}_{p,r}}^{l-1}
+\|U^2_{xx}\|_{B^{s}_{p,r}}^{l-1}
+\|U^1\|_{B^{s}_{p,r}}+\|U^1_x\|_{B^{s}_{p,r}}+\|U^2\|_{B^{s}_{p,r}}+\|U^2_x\|_{B^{s}_{p,r}})\|M^2\|_{B^{s}_{p,r}}\\
\lesssim&\|M^{12}\|_{B^{s-1}_{p,r}}(\|M^1\|_{B^{s}_{p,r}}^{q}+\|M^2\|_{B^{s}_{p,r}}^{q}+1
),
\end{align*}
with $q=\max\{l,2\}.$

Thus, for the case (1) $s>\max\{1-\frac{1}{p},\frac{1}{p}\}$ and $s\neq 2+\frac{1}{p}$, using Lemma \ref{gai} with the above three inequalities and $p_1=p$ and
\begin{align*}
  V_{p_1}'(t)&=\|\partial_xH^1\|_{B^{s-2}_{p,r}}+\|\partial_xH^1\|_{B^{\frac{1}{p}}_{p,r}\cap L^\infty}\leq\|\partial_xH^1\|_{B^{s}_{p,r}},
\end{align*}
we have
\begin{align}\label{shou}
  \|M^{12}(t)\|_{B^{s-1}_{p,r}}\leq & \|M^{12}_0\|_{B^{s-1}_{p,r}}+\int_0^t\|(F_1+F_2+F_3)(\tau)\|_{B^{s-1}_{p,r}}d\tau+C\int_0^t V_{p_1}'(\tau)\|M^{12}(\tau)\|_{B^{s-1}_{p,r}}d\tau\\\nonumber\leq & \|M^{12}_0\|_{B^{s-1}_{p,r}}+C\int_0^t
  \|M^{12}(\tau)\|_{B^{s-1}_{p,r}}(\|M^1(\tau)\|_{B^{s}_{p,r}}^{q}+\|M^2(\tau)\|_{B^{s}_{p,r}}^{q}+1)d\tau
  \\\nonumber&+C\int_0^t \|\partial_xH^1\|_{B^{s}_{p,r}}\|M^{12}(\tau)\|_{B^{s-1}_{p,r}}d\tau\\
  \nonumber\leq & \|M^{12}_0\|_{B^{s-1}_{p,r}}+C\int_0^t
  \|M^{12}(\tau)\|_{B^{s-1}_{p,r}}(\|M^1(\tau)\|_{B^{s}_{p,r}}^{q}+\|M^2(\tau)\|_{B^{s}_{p,r}}^{q}+1)d\tau
  \\\nonumber&+C\int_0^t (\|U^1\|_{B^{s}_{p,r}}^l+\|U^1_x\|_{B^{s}_{p,r}}^l+\|U^1_{xx}\|_{B^{s}_{p,r}}^l)\|M^{12}(\tau)\|_{B^{s-1}_{p,r}}d\tau\\
  \nonumber\leq & \|M^{12}_0\|_{B^{s-1}_{p,r}}+C\int_0^t
  \|M^{12}(\tau)\|_{B^{s-1}_{p,r}}(\|M^1(\tau)\|_{B^{s}_{p,r}}^{q}+\|M^2(\tau)\|_{B^{s}_{p,r}}^{q}+1)d\tau
  .
\end{align}
Hence, the Gronwall lemma gives the inequality (\ref{12}).

For the critical case $(2)$ $s=2+\frac{1}{p}$, let us choose $s_1\in (\max\{1-\frac{1}{p},\frac{1}{p}\}-1, s-1), s_2\in (s-1,s).$ Then $s-1=\theta s_1+(1-\theta)s_2$ with $\theta=\frac{s_2-(s-1)}{s_2-s_1}\in (0,1).$ By using the interpolation inequality and the consequence of the case $(1)$, we get
\begin{align*}
  \|M^{12}(t)\|_{B^{s-1}_{p,r}}\leq & \|M^{12}(t)\|_{B^{s_1}_{p,r}}^\theta\|M^{12}(t)\|_{B^{s_2}_{p,r}}^{1-\theta}\\
  \leq&\big(\|M^{12}_0\|_{B^{s_1}_{p,r}}e^{C\int_0^t(\|M^1(\tau)\|_{B^{s_1+1}_{p,r}}^q
    +\|M^2(\tau)\|_{B^{s_1+1}_{p,r}}^q+1)d\tau}\big)^\theta(
    \|M^{1}(t)\|_{B^{s_2}_{p,r}}+\|M^{2}(t)\|_{B^{s_2}_{p,r}})^{1-\theta}\\
   \leq& \|M^{12}_0\|_{B^{s-1}_{p,r}}^\theta (\|M^{1}(t)\|_{B^s_{p,r}}+\|M^{2}(t)\|_{B^s_{p,r}})^{1-\theta}e^{\theta C\int_0^t(\|M^1(\tau)\|_{B^s_{p,r}}^q
    +\|M^2(\tau)\|_{B^s_{p,r}}^q+1)d\tau},
\end{align*}
which completes the proof of the lemma.\qed

\noindent{Proof of Theorem \ref{pose}.} Since uniqueness in Theorem \ref{pose} is a straightforward corollary of Lemma \ref{weiyi}, we need only to prove the existence of a solution to Eq.(\ref{new}). We shall proceed as follows.\\
First step: constructing approximate solutions.\\
Starting from $M^0=M_0$ we define by induction a sequence $(M^n)_{n\in\mathbb{N}}$ by solving the following linear transport equation
\begin{align}\label{jin}
\left\{
\begin{array}{l}
M^{n+1}_t-H^n M^{n+1}_x=A^n M^n+B^n M^n,\\
M^{n+1}_{|t=0}=M_0,
\end{array}
\right.
\end{align}
where $M^n=(m_1^n,\cdots,m_N^n,n_1^n,\cdots,n_N^n)^T,$ $U^n=(u_1^n,\cdots,u_N^n,v_1^n,\cdots,v_N^n)^T$, $H^n=H(U^n,U_x^n)$, $A^n=A(H^n,H_x^n)$ and $B^n=B(U^n,U_x^n)$.\\
Second step: uniform bounds.\\
Let $q=\max\{l,2\}.$ The condition $s>\max\{1-\frac{1}{p},\frac{1}{p}\}$ yields that $B^{s}_{p,r}$ is an algebra. Thus, we have
\begin{align*}
  &\|A^n M^n+B^n M^n\|_{B^{s}_{p,r}}\leq (\|A^n\|_{B^{s}_{p,r}}+\|B^n\|_{B^{s}_{p,r}})\|M^n\|_{B^{s}_{p,r}}\\
  \lesssim &(1+\|U^n\|_{B^{s}_{p,r}}^l+\|U^n_x\|_{B^{s}_{p,r}}^l+\|U^n_{xx}\|_{B^{s}_{p,r}}^l+
  \|U^n\|_{B^{s}_{p,r}}^2+\|U^n_x\|_{B^{s}_{p,r}}^2)\|M^n\|_{B^{s}_{p,r}}\\
  \lesssim&(1+\|M^n\|_{B^{s}_{p,r}}^q)\|M^n\|_{B^{s}_{p,r}}.
\end{align*}
According to Lemma \ref{gai} with the above inequality and
\begin{align*}
\left\{
\begin{array}{ll}
p_1=p,&~\textit{if}~\max\{1-\frac{1}{p},\frac{1}{p}\}<s,s\neq 1+\frac{1}{p},\\
p_1=\infty,&~\textit{if}~\max\{1-\frac{1}{p},\frac{1}{p}\}<s,s=1+\frac{1}{p}, (\textit{which implies}~ p\neq\infty)
\end{array}
\right.
\end{align*}
and
\begin{align*}
V_{p_1}'(t)&=\left\{
\begin{array}{ll}
\|\partial_xH^n\|_{B^{s-1}_{p,r}}+\|\partial_xH^n\|_{B^{\frac{1}{p}}_{p,r}\cap L^\infty},&~\textit{if}~\max\{1-\frac{1}{p},\frac{1}{p}\}<s,s\neq 1+\frac{1}{p},\\
\|\partial_xH^n\|_{B^{s-1}_{\infty,r}}\lesssim\|\partial_xH^n\|_{B^{s-1+\frac{1}{p}}_{p,r}},
&~\textit{if}~\max\{1-\frac{1}{p},\frac{1}{p}\}<s,s=1+\frac{1}{p},
\end{array}
\right.
\\&\lesssim \left\{\begin{array}{ll}
\|\partial_xH^n\|_{B^{s}_{p,r}},&~\textit{if}~\max\{1-\frac{1}{p},\frac{1}{p}\}<s,s\neq 1+\frac{1}{p},\\
\|\partial_xH^n\|_{B^{s}_{p,r}},&~\textit{if}~\max\{1-\frac{1}{p},\frac{1}{p}\}<s,s=1+\frac{1}{p},
\end{array}
\right.
\end{align*}
we get
  \begin{align*}
  \|M^{n+1}(t)\|_{B^s_{p,r}}\leq &\|M_0\|_{B^s_{p,r}}+C\int_0^t(1+\|M^n(\tau)\|_{B^{s}_{p,r}}^q)\|M^n(\tau)\|_{B^{s}_{p,r}}d\tau\\&+C\int_0^t \|\partial_xH^n(\tau)\|_{B^{s}_{p,r}}\|M^{n+1}(\tau)\|_{B^s_{p,r}}d\tau\\\leq &\|M_0\|_{B^s_{p,r}}+C\int_0^t(1+\|M^n(\tau)\|_{B^{s}_{p,r}}^q)\|M^n(\tau)\|_{B^{s}_{p,r}}d\tau\\&+C\int_0^t (\|U^n\|_{B^{s}_{p,r}}^l+\|U^n_x\|_{B^{s}_{p,r}}^l+\|U^n_{xx}\|_{B^{s}_{p,r}}^l)\|M^{n+1}(\tau)\|_{B^s_{p,r}}d\tau\\\leq &\|M_0\|_{B^s_{p,r}}+C\int_0^t(1+\|M^n(\tau)\|_{B^{s}_{p,r}}^q)\|M^n(\tau)\|_{B^{s}_{p,r}}d\tau\\&+C\int_0^t (\|M^n(\tau)\|_{B^{s}_{p,r}}^q+1)\|M^{n+1}(\tau)\|_{B^s_{p,r}}d\tau.
\end{align*}
The Gronwall lemma yields that
\begin{align}\label{youjie2}
  \|M^{n+1}(t)\|_{B^s_{p,r}}\leq &e^{C\int_0^t(1+\|M^n(\tau)\|_{B^{s}_{p,r}}^q)d\tau}\Big(\|M_0\|_{B^s_{p,r}}\\\nonumber
  &+C\int_0^te^{-C\int_0^\tau(1+\|M^n(t')\|_{B^{s}_{p,r}}^q)dt'}
  (1+\|M^n(\tau)\|_{B^{s}_{p,r}}^q)
  \|M^n(\tau)\|_{B^{s}_{p,r}}d\tau\Big).
\end{align}
Notice that $f(t)=\frac{f_0e^{2Ct}}{(1+f_0^q-f_0^qe^{2Ctq})^{\frac{1}{q}}}$ is the solution to the following equation:
\begin{align}\label{fff}
  f(t)= &e^{C\int_0^t(1+f^q(\tau))d\tau}\Big(f_0
  +C\int_0^te^{-C\int_0^\tau(1+f^q(t')dt'}
  (1+f^q(\tau))f(\tau)d\tau\Big).
\end{align}
We fix a $T>0$ such that $1+\|M_0\|_{B^s_{p,r}}^q-\|M_0\|_{B^s_{p,r}}^qe^{2qCT}>0$ and suppose that
\begin{align*}
  \forall~t\in[0,T],~~\|M^n\|_{B^s_{p,r}}\leq \frac{\|M_0\|_{B^s_{p,r}}e^{2Ct}}{(1+\|M_0\|_{B^s_{p,r}}^q-\|M_0\|_{B^s_{p,r}}^qe^{2Ctq})^{\frac{1}{q}}}.
\end{align*}
Plugging the above inequality into (\ref{youjie2}) and using (\ref{fff}) yield
\begin{align*}
  \|M^{n+1}(t)\|_{B^s_{p,r}}\leq \frac{\|M_0\|_{B^s_{p,r}}e^{2Ct}}{(1+\|M_0\|_{B^s_{p,r}}^q-\|M_0\|_{B^s_{p,r}}^qe^{2Ctq})^{\frac{1}{q}}}.
\end{align*}
Therefore, $(M^n)_{n\in\mathbb{N}}$ is bounded in $L^\infty(0,T;B^s_{p,r}).$\\
Third step: convergence.\\
Similar to the proof of (\ref{shou}), we have, for $s>\max\{1-\frac{1}{p},\frac{1}{p}\}$ and $s\neq 2+\frac{1}{p}$,
\begin{align*}
  &\|(M^{n+m+1}-M^{n+1})(t)\|_{B^{s-1}_{p,r}}\\\leq &C\int_0^t
  \|(M^{n+m}-M^{n})(\tau)\|_{B^{s-1}_{p,r}}
  (\|M^{n}(\tau)\|_{B^{s}_{p,r}}^q+\|M^{n+1}(\tau)\|_{B^{s}_{p,r}}^q
  +\|M^{n+m}(\tau)\|_{B^{s}_{p,r}}^q+1)d\tau\\&+C\int_0^t
  \|(M^{n+m+1}-M^{n+1})(\tau)\|_{B^{s-1}_{p,r}}
  (\|M^{n+m}(\tau)\|_{B^{s}_{p,r}}^q+1)d\tau.
\end{align*}
Taking advantage of the Gronwall inequality gives
\begin{align*}
  &\|(M^{n+m+1}-M^{n+1})(t)\|_{B^{s-1}_{p,r}}\\\leq &Ce^{C\int_0^t(\|M^{n+m}(t')\|_{B^{s}_{p,r}}^q+1)dt'}\int_0^t
  e^{-C\int_0^\tau(\|M^{n+m}(t')\|_{B^{s}_{p,r}}^q+1)dt'}\|(M^{n+m}-M^{n})(\tau)\|_{B^{s-1}_{p,r}}\\
  &\times(\|M^{n}(\tau)\|_{B^{s}_{p,r}}^q+\|M^{n+1}(\tau)\|_{B^{s}_{p,r}}^q
  +\|M^{n+m}(\tau)\|_{B^{s}_{p,r}}^q+1)d\tau.
\end{align*}
Since $(M^n)_{n\in\mathbb{N}}$ is bounded in $L^\infty(0,T;B^s_{p,r}),$ we finally get a constant $C_T$, independent of $n$ and $m$, such that
\begin{align*}
  \|(M^{n+m+1}-M^{n+1})(t)\|_{B^{s-1}_{p,r}}\leq C_T\int_0^t
  \|(M^{n+m}-M^{n})(\tau)\|_{B^{s-1}_{p,r}}d\tau.
\end{align*}
Finally, arguing by induction, we arrive at
\begin{align*}
  \|(M^{n+m+1}-M^{n+1})(t)\|_{B^{s-1}_{p,r}}\leq \frac{(TC_T)^{n+1}}{(n+1)!}\|M^{m}-M^{0}\|_{L^\infty_T(B^{s-1}_{p,r})}\leq C_T\frac{(TC_T)^{n+1}}{(n+1)!},
  \end{align*}
which implies that $(M^n)_{n\in\mathbb{N}}$ is a Cauchy sequence in $L^\infty(0,T;B^{s-1}_{p,r}).$

For the critical case $s=2+\frac{1}{p}$, from the above argument, we get that $(M^n)_{n\in\mathbb{N}}$ is a Cauchy sequence in $L^\infty(0,T;B^{s-1-\varepsilon}_{p,r})$ with sufficiently small $\varepsilon$. Then applying the interpolation method with uniform bounds in $L^\infty(0,T;B^s_{p,r})$ obtained in the second step, we show that $(M^n)_{n\in\mathbb{N}}$ is also a Cauchy sequence in $L^\infty(0,T;B^{s-1}_{p,r})$ for the critical case.\\
Final step: conclusion.\\
Let $M$ be the limit of the sequence $(M^n)_{n\in\mathbb{N}}$ in $L^\infty(0,T;B^{s-1}_{p,r})$. According to the Fatou lemma \ref{Fatou}, $M$ also belongs to $L^\infty(0,T;B^s_{p,r})$. It is then easy to pass to the limit in Eq.(\ref{jin}) and to conclude that $M$ is a solution of Eq.(\ref{new}). Note that $A(U,U_x)M+B(H,H_x)M$ of Eq.(\ref{new}) also belongs to $L^\infty(0,T;B^s_{p,r})$. According to Lemma \ref{gai}, we have $M\in C([0,T];B^s_{p,r})$  if $r<\infty,$ or $M\in\Big(\bigcap_{s'<s}C([0,T];B^{s'}_{p,r})\Big)$, if $r=\infty.$ Again using the equation, we see that $M_t\in C([0,T];B^{s-1}_{p,r})$  if $r<\infty,$ or $M_t\in\Big(\bigcap_{s'<s}C([0,T];B^{s'-1}_{p,r})\Big)$, if $r=\infty.$
This completes the proof of Theorem \ref{pose}.\qed.
\subsection{A continuation criterion}
In this subsection, we state a continuation criterion for Eq.(\ref{new}).
\begin{theo}\label{baopo1}
  Let $M_0\in  B^s_{p,r}$ with $1\leq p,r\leq \infty$, $s>\max\{1-\frac{1}{p},\frac{1}{p}\}$ and $T>0$ be the maximal existence time of the corresponding solution $M$ to Eq.(\ref{new}). If $T$ is finite, then we have
  \begin{align*}
    \int_0^T\|M(\tau)\|_{L^\infty}^qd\tau=\infty,
  \end{align*}where $q=\max\{l,2\}$ ($l$ is the polynomial order of $H$).
  \end{theo}
  \noindent{Proof.} 
  For any $0<\sigma\leq s$, applying Proposition \ref{chengji} (i), we have \begin{align}\label{baopoyong}
    &\|A(H,H_x)M+B(U,U_x)M\|_{B^\sigma_{p,r}}\\\nonumber \leq & (\|A(H,H_x)\|_{L^\infty}+\|B(U,U_x)\|_{L^\infty})\|M\|_{B^\sigma_{p,r}}+
    (\|A(H,H_x)\|_{B^\sigma_{p,r}}+\|B(U,U_x)\|_{B^\sigma_{p,r}})\|M\|_{L^\infty}\\\nonumber\leq
    &(\|U\|_{L^\infty}^l+\|U_x\|_{L^\infty}^l+\|U_{xx}\|_{L^\infty}^l+1+
    \|U\|_{L^\infty}^2+\|U_x\|_{L^\infty}^2)\|M\|_{B^\sigma_{p,r}}\\\nonumber
    &+\Big((\|U\|_{B^\sigma_{p,r}}+\|U_x\|_{B^\sigma_{p,r}}+\|U_{xx}\|_{B^\sigma_{p,r}}+1)
    (\|U\|_{L^\infty}^{l-1}+\|U_x\|_{L^\infty}^{l-1}+\|U_{xx}\|_{L^\infty}^{l-1}+1)\\\nonumber&+
    (\|U\|_{B^\sigma_{p,r}}+\|U_x\|_{B^\sigma_{p,r}})(\|U\|_{L^\infty}+\|U_x\|_{L^\infty})\Big)
    \|M\|_{L^\infty}\\\nonumber\leq &
    (\|M\|_{L^\infty}^q+1)(\|M\|_{B^\sigma_{p,r}}+1).
  \end{align}
  We now consider the case $1< p<\infty$.\\
  Step 1. If $\sigma>1,$ then we claim that \begin{align}\label{s>}
    \|M\|_{L^\infty_T(B^{1}_{p,r})}<\infty,~ \textit{and}~\int_0^T\|M(\tau)\|_{L^\infty}^qd\tau<\infty\Rightarrow\|M\|_{L^\infty_T(B^\sigma_{p,r})}<\infty.
  \end{align}
  In fact, by using (\ref{baopoyong}) and Lemma \ref{gai} with $p_1=\infty$ and
  \begin{align*}
  V'_{p_1}(t)&=\|\partial_x H\|_{B^{\sigma-1}_{\infty,r}}\leq \|\partial_x H\|_{B^{\sigma-1+\frac{1}{p}}_{p,r}}\\&\leq  \|U\|_{B^{\sigma-1+\frac{1}{p}}_{p,r}}^{l}
  +\|U_x\|_{B^{\sigma-1+\frac{1}{p}}_{p,r}}^{l}
  +\|U_{xx}\|_{B^{\sigma-1+\frac{1}{p}}_{p,r}}^{l}
  \leq \|M\|_{B^{\sigma-1+\frac{1}{p}}_{p,r}}^l,
  \end{align*}
    we have
  \begin{align*}
  \|M(t)\|_{B^\sigma_{p,r}}\leq &\|M_0\|_{B^\sigma_{p,r}}+C\int_0^t(\|M(\tau)\|_{L^\infty}^q+1)(\|M(\tau)\|_{B^\sigma_{p,r}}
  +1)d\tau\\&+C\int_0^t \|M\|_{B^{\sigma-1+\frac{1}{p}}_{p,r}}^l\|M(\tau)\|_{B^\sigma_{p,r}}d\tau.
\end{align*}
Hence, the Gronwall lemma gives
\begin{align*}
   \|M(t)\|_{B^\sigma_{p,r}}+1\leq (\|M_0\|_{B^\sigma_{p,r}}+1)e^{C\int_0^t(\|M(\tau)\|_{L^\infty}^q+1
   +\|M\|_{B^{\sigma-1+\frac{1}{p}}_{p,r}}^l)
  d\tau},
\end{align*}
which implies
\begin{align*}
    \|M\|_{L^\infty_T(B^{\sigma-1+\frac{1}{p}}_{p,r})}<\infty,~\textit{and}~ \int_0^T\|M(\tau)\|_{L^\infty}^qd\tau<\infty\Rightarrow\|M\|_{L^\infty_T(B^\sigma_{p,r})}<\infty
    .
  \end{align*}
 If $\sigma-1+\frac{1}{p}>1,$ then repeat the above process. Clearly, this process stops within a finite number of steps. Our claim (\ref{s>}) is guaranteed.\\
 Step 2. If $\sigma=1,$  then by using (\ref{baopoyong}) and Lemma \ref{gai} with $p_1=p$ and
  \begin{align*}
  V'_{p_1}(t)=&\|\partial_x H\|_{B^{\frac{1}{p}}_{p,\infty}\cap L^{\infty}} \leq \|U\|_{B^{\frac{1}{p}}_{p,\infty}\cap L^\infty}^{l}+\|U_x\|_{B^{\frac{1}{p}}_{p,\infty}\cap L^\infty}^{l}+\|U_{xx}\|_{B^{\frac{1}{p}}_{p,\infty}\cap L^\infty}^{l}\\
  \leq& \|M\|_{B^{\frac{1}{p}}_{p,\infty}\cap L^\infty}^l\lesssim \|M\|_{B^{\frac{1}{p}}_{p,\infty}}^l+ \|M\|_{L^\infty}^q+1,
  \end{align*}
    we have
  \begin{align*}
  \|M(t)\|_{B^\sigma_{p,r}}\leq &\|M_0\|_{B^\sigma_{p,r}}+C\int_0^t(\|M(\tau)\|_{L^\infty}^q+1)(\|M(\tau)\|_{B^\sigma_{p,r}}
  +1)d\tau\\&+C\int_0^t (\|M\|_{B^{\frac{1}{p}}_{p,r}}^l+ \|M\|_{L^\infty}^q+1)\|M(\tau)\|_{B^\sigma_{p,r}}d\tau.
\end{align*}
Hence, the Gronwall lemma gives
\begin{align*}
   \|M(t)\|_{B^\sigma_{p,r}}+1\leq (\|M_0\|_{B^\sigma_{p,r}}+1)e^{C\int_0^t(\|M(\tau)\|_{L^\infty}^q+1
   +\|M\|_{B^{\frac{1}{p}}_{p,r}}^l)
  d\tau},
\end{align*}
which implies
 \begin{align}\label{s=}
    \|M\|_{L^\infty_T(B^\frac{1}{p}_{p,r})} <\infty~\textit{and}~ \int_0^T\|M(\tau)\|_{L^\infty}^qd\tau<\infty\Rightarrow
    \|M\|_{L^\infty_T(B^\sigma_{p,r})}<\infty.
  \end{align}
  Step 3.
 If $\sigma\in (0, 1),$ applying Lemma \ref{gai} with $p_1=\infty$ and
  \begin{align*}
  V'_{p_1}(t)&=\|\partial_x H\|_{B^{0}_{\infty,\infty}\cap L^{\infty}}\leq \|\partial_x H\|_{L^{\infty}}\\&\leq \|U\|_{L^\infty}^{l}+\|U_x\|_{L^\infty}^{l}+\|U_{xx}\|_{L^\infty}^{l}
  \leq \|M\|_{L^\infty}^q+1,
  \end{align*}
  we have
  \begin{align*}
  \|M(t)\|_{B^\sigma_{p,r}}\leq &\|M_0\|_{B^\sigma_{p,r}}+C\int_0^t(\|M(\tau)\|_{L^\infty}^q+1)(\|M(\tau)\|_{B^s_{p,r}}
  +1)d\tau\\&+C\int_0^t (\|M\|_{L^\infty}^q+1)\|M(\tau)\|_{B^\sigma_{p,r}}d\tau.
\end{align*}
Hence, the Gronwall lemma gives
\begin{align*}
   \|M(t)\|_{B^\sigma_{p,r}}+1\leq (\|M_0\|_{B^\sigma_{p,r}}+1)e^{C\int_0^t(\|M(\tau)\|_{L^\infty}^q+1)
  d\tau},
\end{align*}
which implies
 \begin{align}\label{s=}
 \int_0^T\|M(\tau)\|_{L^\infty}^qd\tau<\infty
 \Rightarrow\|M\|_{L^\infty_T(B^\sigma_{p,r})}<\infty.
  \end{align}
Therefore, for all $s>\max\{1-\frac{1}{p},\frac{1}{p}\}$,  if $T<\infty,$ and $
    \int_0^T\|M(\tau)\|_{L^\infty}^qd\tau<\infty
$, then we have $\limsup_{t\rightarrow T}$ $\|M(t)\|_{B^s_{p,r}}<\infty.$

The cases $p=1$ and $p=\infty$ can be treated similarly. We also have for $s>1$, if $T<\infty,$ and $
    \int_0^T\|M(\tau)\|_{L^\infty}^qd\tau<\infty,
$ then $\limsup_{t\rightarrow T}\|M(t)\|_{B^s_{p,r}}<\infty.$ For the sake of simplicity, we omit the details here.$^1$

\footnotetext[1]{We present a simple flow chart. For $p=\infty,$ $\sigma>1,$ let $\varepsilon\in (0,1).$ $\|M\|_{B^\sigma_{\infty,r}}\overset{\textit{determined~ by}}{\longrightarrow}\|\partial_x H\|_{B^{\sigma-1}_{\infty,r}}\lesssim \|M\|_{B^{\sigma-1}_{\infty,r}}\overset{\textit{determined~ by}}{\dashrightarrow}\|M\|_{B^{\varepsilon+1}_{\infty,r}}\overset{\textit{determined~ by}}{\longrightarrow}
\|\partial_x H\|_{B^{\varepsilon}_{\infty,r}}\lesssim \|M\|_{B^{\varepsilon}_{\infty,r}}\overset{\textit{determined~ by}}{\longrightarrow}\|\partial_x H\|_{B^{0}_{\infty,\infty}\cap L^{\infty}}\lesssim \|M\|_{L^{\infty}}.$\\
For $p=1$, $\sigma>1,$ choose $p_1$ such that $1<p_1<\infty$ and $\sigma>1+\frac{1}{p_1}.$ $\|M\|_{B^\sigma_{1,r}}\overset{\textit{determined~ by}}{\longrightarrow}\|\partial_x H\|_{B^{\sigma-1}_{p_1,r}}\lesssim \|M\|_{B^{\sigma-1}_{p_1,r}} \overset{\textit{determined~ by}}{\dashrightarrow}       $turn to the case (1) $1<p_1<\infty$ $\overset{\textit{determined~ by}}{\dashrightarrow}\|M\|_{L^{\infty}}$.}

Finally, if $\limsup_{t\rightarrow T}\|M(t)\|_{B^s_{p,r}}<\infty,$ then by Theorem \ref{pose}, we can extent the solution $M$ beyond $T$, which is a contradiction with the assumption of $T$. Then we must have  $
    \int_0^T\|M(\tau)\|_{L^\infty}^qd\tau=\infty.$ This completes the proof of the theorem. \qed

Combining Theorem \ref{pose} and Theorem \ref{baopo1}, we readily obtain the following corollary.
\begin{coro}\label{tui}
  Let $M_0\in  B^s_{p,r}$ with $1\leq p,r\leq \infty$ and $s>\max\{1-\frac{1}{p},\frac{1}{p}\}$ and $T>0$ be the maximal existence time of the corresponding solution $M$ to Eq.(\ref{new}). Then the solution $M$ blows up in finite time if and only if
 $
    \limsup_{t\rightarrow T}\|M(t)\|_{L^\infty}=\infty.
$
\end{coro}
\begin{rema} \label{rreemm}Apparently, for every $s\in \mathbb{R},$ $B^s_{2,2}=H^s.$ Theorem \ref{pose}, Theorem \ref{baopo1} and Corollary \ref{tui} hold true in the corresponding Sobolev spaces $H^s$ with $s>\frac{1}{2}$, which recovers the corresponding results in \cite{yan} and \cite{Gui} as $N=1,~ H=-\frac{1}{2}(u_1v_1-u_{1x}v_{1x})$ and $N=1,~ H=-\frac{1}{2}(u_1v_1-u_{1x}v_{1x}),v=2u$, respectively.
\end{rema}
\begin{rema} We pointed out that if $N=1$ and $H=-\frac{1}{2}(u_1v_1-u_{1x}v_{1x})$, then Theorem \ref{pose} improves the corresponding result in \cite{yan}, where $s>\max\{1-\frac{1}{p},\frac{1}{p},\frac{1}{2}\}$ but $s\neq 1+\frac{1}{p}$. Besides, Theorem \ref{baopo1}, which works in Besov spaces, also improves the corresponding result in \cite{yan}, where the corresponding blow-up scenario works only in Sobolev spaces.
\end{rema}

\section{Global existence and blow-up phenomena for the two-component subsystems}
\subsection{$N=1,H=-\frac{1}{2}(u-u_x)(v+v_x)$}
\subsubsection{A precise blow-up scenario}
As mentioned in the Introduction, for $N=1$ and $H=-\frac{1}{2}(u-u_x)(v+v_x)$, Eq.(\ref{s0}) is reduced to the following system:
\begin{align}\label{1.2}
\left\{
\begin{array}{l}
m_{t}+\frac{1}{2}\big((u-u_x)(v+v_x) m\big)_x=0,\\[1ex]
n_{t}+\frac{1}{2}\big((u-u_x)(v+v_x) n\big)_x=0,\\[1ex]
(m,n)|_{t=0}=(m_{0},n_{0}),
\end{array}
\right.
\end{align}
where $m=u-u_{xx}$ and $n=v-v_{xx}$.

Consider the following initial value problem
\begin{align}\label{q}
\left\{
\begin{array}{l}
q_{t}(t,x)=\frac{1}{2}(u-u_x)(v+v_x)(t,q),~~~t\in[0,T),\\[1ex]
q(0,x)=x, ~~~x\in \mathbb{R}.
\end{array}
\right.
\end{align}

\begin{lemm}\label{qq0}
Let $m_{0},n_{0}\in H^s~(s>\frac{1}{2})$, and let $T>0$ be the maximal existence time of the corresponding solution $(m,n)$ to Eq.(\ref{1.2}). Then Eq.(\ref{q}) has a unique solution $q\in C^1([0,T]\times \mathbb{R};\mathbb{R}).$ Moreover, the mapping $q(t,\cdot)$ $(t\in[0,T))$ is an increasing diffeomorphism of $\mathbb{R},$ with
\begin{align}\label{qx}
q_x(t,x)=exp\big(\int_0^t\frac{1}{2}\big(m(v+v_x)-n(u-u_x)\big)(\tau,q(\tau,x))d\tau\big).
\end{align}
\end{lemm}
\noindent{Proof}. According to Remark \ref{rreemm}, we get that $m,n\in C([0,T];H^s)\cap C^1([0,T];H^{s-1})$ with $s>\frac{1}{2},$ from which we deduce that $\frac{1}{2}(u-u_x)(v+v_x)$ is bounded and Lipschitz continuous in the space variable $x$ and of class $C^1$ in time variable $t$, then the classical ODE theory ensures that Eq.(\ref{q}) has a unique solution $q\in C^1([0,T]\times \mathbb{R};\mathbb{R}).$
Differentiating Eq.(\ref{q})with respect to $x$ gives
\begin{align}\label{q_x}
\left\{
\begin{array}{l}
{q_x}_{t}(t,x)=\frac{1}{2}\big(m(v+v_x)-n(u-u_x)\big)(t,q)q_x(t,x),~~~t\in[0,T),\\[1ex]
q_x(0,x)=1, ~~~x\in \mathbb{R},
\end{array}
\right.
\end{align}
which leads to (\ref{qx}). So, the mapping $q(t,\cdot)$ $(t\in[0,T))$ is an increasing diffeomorphism of $\mathbb{R}.$\qed
\begin{lemm}\label{ll11}
  Let $m_{0},n_{0}\in H^s~(s>\frac{1}{2})$, and let $T>0$ be the maximal existence time of the corresponding solution $(m,n)$ to Eq.(\ref{1.2}). Then, we have for all $t\in[0,T),$
  \begin{align}
    &m(t,q(t,x))q_x(t,x)=m_{0}(x),\label{m1}\\
    &n(t,q(t,x))q_x(t,x)=n_{0}(x).\label{m2}
  \end{align}
\end{lemm}
\noindent{Proof.} Combining Eq.(\ref{q}), Lemma \ref{qq0} and Eq.(\ref{1.2}), we have
\begin{align*}
\frac{d}{dt}\big(m(t,q(t,x))q_x(t,x)\big)=
&\big(m_{t}(t,q)+m_{x}(t,q)q_t(t,x)\big)q_x(t,x)+m(t,q)q_{xt}(t,x)\\
=&\big(m_{t}(t,q)+\frac{1}{2}\big((u-u_x)(v+v_x) m\big)_x(t,q)\big)q_x(t,x)=0
.
\end{align*}
Therefore, the Gronwall inequality yields (\ref{m1}). Similar arguments lead to (\ref{m2}). This completes the proof of the lemma.\qed

The following theorem shows a precise blow-up scenario for Eq.(\ref{1.2}).
\begin{theo}\label{bao222}
   Let $m_{0},n_{0}\in H^s$ $(s>\frac{1}{2})$, and let $T>0$ be the maximal existence time of the corresponding solution $(m,n)$ to Eq.(\ref{1.2}). Then the solution $(m,n)$ blows up in finite time if and only if
   \begin{align*}
     \liminf_{t\rightarrow T}\inf_{x\in \mathbb{R}}\big(m(v+v_x)-n(u-u_x)\big)(t,x)=-\infty.
   \end{align*}
\end{theo}
\noindent{Proof.} Assume that the solution $(m,n)$ blows up in finite time $T$ and there exists a constant $C$ such that
\begin{align*}
  \big(m(v+v_x)-n(u-u_x)\big)(t,x)\geq -C,~~\forall~(t,x)\in [0,T)\times \mathbb{R}.
\end{align*}
By (\ref{qx}) and Lemma \ref{ll11}, we have that
\begin{align*}
  \|m(t)\|_{L^\infty}+\|n(t)\|_{L^\infty}\leq (\|m_{0}\|_{L^\infty}+\|n_{0}\|_{L^\infty})e^{Ct}, ~~\forall~t\in [0,T),
\end{align*}
which contradicts to Corollary \ref{tui}.

On the other hand, if $\liminf_{t\rightarrow T}\inf_{x\in \mathbb{R}}\big(m(v+v_x)-n(u-u_x)\big)(t,x)=-\infty
  $, then we can get $$\limsup_{t\rightarrow T} \|m(t)\|_{L^\infty}=\infty~ \textit{or}~~\limsup_{t\rightarrow T}\|n(t)\|_{L^\infty} =\infty.$$ Thus according to Corollary \ref{tui}, the solution $(m,n)$ blows up. This completes the proof of the theorem.\qed
\subsubsection{Global existence}
We now give a global existence result.
\begin{theo}\label{l1.1}
Let $m_{0},n_{0}\in H^s$ $(s>\frac{1}{2})$. Assume that $supp~{m_{0}}\in[b,\infty),~supp~{n_{0}}\in(-\infty,a],$ with $a\leq b.$ Then the corresponding solution $(m,n)$ to Eq.(\ref{1.2})
 exists globally in time.
\end{theo}
\noindent{Proof.} Note that, according to Lemma \ref{qq0}, the function $q(t,x)$ is an increasing diffeomorphism of $\mathbb{R}$ with $q_x(t,x)>0$ with respect to time $t$. Thus $a\leq b$ implies $q(t,a)\leq q(t,b).$ We infer from Lemma \ref{qq0} and Lemma \ref{ll11} that for all $t\in[0,T),$ we have
\begin{align}\label{gg}
\left\{
\begin{array}{l}
  m(t,x)=0,~\textit{if}~~x<q(t,b),\\
  n(t,x)=0,~\textit{if}~~x>q(t,a).
\end{array}
\right.
\end{align}
Noticing
\begin{align*}
u(t,x)-u_{x}(t,x)=e^{-x}\int_{-\infty}^xe^ym(t,y)dy,\\
v(t,x)+v_{x}(t,x)=e^x\int_x^\infty e^{-y}n(t,y)dy,
\end{align*}
we have
\begin{align}\label{g}
\left\{
\begin{array}{l}
  u(t,x)-u_{x}(t,x)=0,~\textit{if}~~x\leq q(t,b),\\
  v(t,x)+v_{x}(t,x)=0,~\textit{if}~~x\geq q(t,a).
\end{array}
\right.
\end{align}
Therefore, for Eq.(\ref{1.2}), $\big(m(v+v_x)-n(u-u_x)\big)(t,x)=0$ on $\mathbb{R}$ for all $t\in[0,T)$.
Then Theorem \ref{bao222} implies $T=\infty.$ This proves the solution $(m,n)$ exists globally in time.\qed
\subsubsection{Blow-up phenomena}
As a straight corollary of Lemma \ref{qq0}- \ref{ll11}, we have the following lemma.
\begin{lemm}\label{LL1}
Let $m_{0},n_{0}\in H^s~(s>\frac{1}{2})$, and let $T>0$ be the maximal existence time of the corresponding solution $(m,n)$ to Eq.(\ref{1.2}). Assume further $m_0,n_0\in L^1$. Then we have for all $t\in[0,T),$
\begin{align*}
  \|m(t)\|_{L^1}=\|m_0\|_{L^1},\|n(t)\|_{L^1}=\|n_0\|_{L^1}.
\end{align*}
\end{lemm}
Now we derive two useful conservation laws for Eq.(\ref{1.2}).
\begin{lemm}\label{L0}
Let $m_{0},n_{0}\in H^s$ with $s>\frac{1}{2},$ and let $T>0$ be the maximal existence time of the corresponding solution $(m,n)$ to Eq.(\ref{1.2}). Then we have that for all $t\in[0,T),$
\begin{align*}
  \int_{R}m(v+v_x)(t,x)dx=\int_{R}m_{0}(v_0+v_{0x})dx,\\
 \int_{R}n(u-u_x)(t,x)dx=\int_{R}n_{0}(u_0-u_{0x})dx.
\end{align*}
\end{lemm}
\noindent{Proof}. By Eq.(\ref{1.2}), we have
\begin{align*}
  &\frac{d}{dt}\int_{R}m(v+v_x)(t,x)dx=\frac{d}{dt}\int_{R}n(u-u_x)(t,x)dx\\
  =&\int_{\mathbb{R}}\big((v+v_x)m_{t}+(u-u_x)n_{t}\big)(t,x)dx\\
  =&\frac{1}{2}\int_{\mathbb{R}}(u-u_x)(v+v_x)\big(m(v_x+v_{xx})+n(u_x-u_{xx}\big)(t,x)dx\\
  =&\frac{1}{2}\int_{\mathbb{R}}(u-u_x)(v+v_x)\big(m(v+v_x)-n(u-u_x)\big)(t,x)dx\\
  =&\frac{1}{2}\int_{\mathbb{R}}(u-u_x)(v+v_x)\partial_x\big((u-u_x)(v+v_x)\big)(t,x)dx=0.
\end{align*}
This completes the proof of the lemma.\qed
\begin{lemm}\label{ll1.1}
Let $m_{0},n_{0}\in H^s$ $(s>\frac{1}{2})$, and let $T>0$ be the maximal existence time of the corresponding solution $(m,n)$ to Eq.(\ref{1.2}).
 Assume that $m_{0}$ and $ n_{0}$ do not change sign. Then there exists a constant $C=C(\|(v_0+v_{0x})m_{0}\|_{L^1},\|(u_0-u_{0x})n_{0}\|_{L^1},\|u_{0}\|_{H^1},\|v_{0}\|_{H^1})$ such that
\begin{align}
|u_{x}(t,x)|\leq |u_{}(t,x)|,~~~|v_{x}(t,x)|\leq |v_{}(t,x)|,\\
  \|u(t)\|_{H^1}+\|v(t)\|_{H^1}\leq Ce^{Ct},~\forall t\in[0,T).\label{u11}
  \end{align}
\end{lemm}
\noindent{Proof}. One can assume without loss of generality that $m_{0}\geq 0,~n_{0}\geq 0$ for all $x\in \mathbb{R}.$ Since $m_{0}\geq 0,$ (\ref{qx}) and (\ref{m1}) imply that
\begin{align}\label{m11}
m(t,x)\geq 0, ~~~\forall (t,x)\in [0,T)\times \mathbb{R}.
\end{align}
Noticing
\begin{align*}
  u(t,x)=(1-\partial_x^2)^{-1}m(t,x)=\frac{1}{2}\int_\mathbb{R}e^{-|x-y|}m(t,y)dy,
\end{align*}
we obtain
\begin{align*}
  u(t,x)&=\frac{e^{-x}}{2}\int_{-\infty}^xe^ym(t,y)dy+\frac{e^x}{2}\int_x^\infty e^{-y}m(t,y)dy,\end{align*} and \begin{align*}
  u_{x}(t,x)&=-\frac{e^{-x}}{2}\int_{-\infty}^xe^ym(t,y)dy+\frac{e^x}{2}\int_x^\infty e^{-y}m(t,y)dy,
\end{align*}
which lead to
\begin{align}
  u(t,x)+ u_{x}(t,x)&=e^x\int_x^\infty e^{-y}m(t,y)dy\geq 0,\label{f11}\\
  u(t,x)-u_{x}(t,x)&=e^{-x}\int_{-\infty}^xe^ym(t,y)dy\geq 0.
\end{align}
From the above two inequalities, we have
\begin{align}\label{u1}
|u_{x}(t,x)|\leq u(t,x), ~~~\forall (t,x)\in [0,T)\times \mathbb{R}.
\end{align}
Similar arguments lead to
\begin{align}
  &n(t,x)\geq 0, ~~~\forall (t,x)\in [0,T)\times \mathbb{R}.\label{m22}\\
  &v(t,x)+v_{x}(t,x)=e^{x}\int_{-\infty}^xe^{-y}n(t,y)dy\geq 0,\label{f22}\\
  &v(t,x)-v_{x}(t,x)=e^{-x}\int_{-\infty}^xe^yn(t,y)dy\geq 0,\\
  &|v_{x}(t,x)|\leq v(t,x), ~~~\forall (t,x)\in [0,T)\times \mathbb{R}.\label{u2}
\end{align}
Using Eq.(\ref{1.2}), we get
\begin{align*}
  &\frac{1}{2}\frac{d}{dt}(\|u(t)\|_{H^1}^2+\|v(t)\|_{H^1}^2)=
  \int_{\mathbb{R}}(m_{t}u+n_{t}v)(t,x)dx\\
  =&\frac{1}{2}\int_{\mathbb{R}}\big((u-u_x)(v+v_x)mu_{x}+(u-u_x)(v+v_x)nv_{x}\big)(t,x)dx\\
  \leq&\frac{1}{2}(\|\big((u-u_x)u_{x}\big)(t)\|_{L^\infty}\|\big((v+v_x)m\big)(t)\|_{L^1}
  +\|\big((v+v_x)v_{x}\big)(t)\|_{L^\infty}\|\big((u-u_x)n\big)(t)\|_{L^1}).
\end{align*}
Using (\ref{u1}) and (\ref{u2}), it yields that
\begin{align*}
  \|\big((u-u_x)u_{x}\big)(t)\|_{L^\infty}
  &\leq 2\|u_{}(t)\|_{L^\infty}^2\leq \|u_{}(t)\|_{H^1}^2,\\
  \|\big((v+v_x)v_{x}\big)(t)\|_{L^\infty}
  &\leq 2\|v_{}(t)\|_{L^\infty}^2\leq \|v_{}(t)\|_{H^1}^2.
\end{align*}
Using Lemma \ref{L0} with the fact that $m,n,u-u_x,v+v_x\geq 0$, we obtain
\begin{align*}
  &\|\big((v+v_x)m\big)(t)\|_{L^1}=\|(v_0+v_{0x})m_0\|_{L^1},\\
 &\|\big((u-u_x)n\big)(t)\|_{L^1}=\|(u_0-u_{0x})n_0\|_{L^1},
\end{align*}
Combining the above three relations, we deduce that
\begin{align*}
  \frac{d}{dt}(\|u(t)\|_{H^1}^2+\|v(t)\|_{H^1}^2)\leq \frac{1}{2}(\|(v_0+v_{0x})m_0\|_{L^1}+\|(u_0-u_{0x})n_0\|_{L^1})(\|u(t)\|_{H^1}^2+\|v(t)\|_{H^1}^2).
\end{align*}
Gronwall's inequality then yields the desired inequality (\ref{u11}). This completes the proof of the lemma.\qed
\begin{lemm}\label{xinjiaL1}
Let $m_{0},n_{0}\in H^s~(s>\frac{1}{2})$, and let $T>0$ be the maximal existence time of the corresponding solution $(m,n)$ to Eq.(\ref{1.2}). Assume further $m_0,n_0\in L^1$. Set $Q(t,x)=\frac{1}{2}(u-u_x)(v+v_x)(t,x).$ Then there exists a constant $C=C(\|m_0\|_{L^1},\|n_0\|_{L^1})$ such that for all $t\in[0,T),$
\begin{align}
&Q_{xt}(t,x)+ \big(Q(Q_{x})_x\big)(t,x)+Q_{x}^2(t,x)\leq C(|m|+|n|)(t,x).
\end{align}
\end{lemm}
\noindent{Proof.}
It is easy to deduce from Eq.(\ref{1.2}) that
\begin{align}\label{a}
&Q_{xt}(t,x)+ \big(Q(Q_{x})_x\big)(t,x)+Q_{x}^2(t,x)\\\nonumber=
&[-(1-\partial_x^2)^{-1}\big((Q_xv)+
\partial_x(Q_xv_{x})\big)m
-(1-\partial_x^2)^{-1}\big(\partial_x(Q_xv)+
(Q_xv_{x})\big)m\\\nonumber
&+(1-\partial_x^2)^{-1}\big((Q_xu)+
\partial_x(Q_xu_{x})\big)n
-(1-\partial_x^2)^{-1}\big(\partial_x(Q_xu)+
(Q_xu_{x})\big)n](t,x),
\end{align}
where $Q_x=\frac{1}{2}\big(m(v+v_x)-n(u-u_x)\big).$
Applying Lemma \ref{LL1}, we arrive at
\begin{align*}
  &(1-\partial_x^2)^{-1}\big((Q_xv)+
\partial_x(Q_xv_{x})\big)(t,x)m(t,x)\\\leq &\|(1-\partial_x^2)^{-1}\big((Q_xv)+
\partial_x(Q_xv_{x})\big)(t)\|_{L^\infty}|m(t,x)|\\\leq & \|\frac{1}{2}e^{-|x|}\|_{L^\infty}(\|Q_x(t)v(t)\|_{L^1}+\|Q_x(t)v_x(t)\|_{L^1}\big)|m(t,x)|
\\\leq&
C\|Q_x(t)\|_{L^1}(\|v(t)\|_{L^\infty}+\|v_x(t)\|_{L^\infty})|m(t,x)|\\\leq & C(\|m(t)\|_{L^1}+\|n(t)\|_{L^1})(\|(u(t)-u_x(t))\|_{L^\infty}+\|(v(t)+v_x(t))\|_{L^\infty})
(\|v(t)\|_{L^\infty}+\|v_x(t)\|_{L^\infty})|m(t,x)|\\
\leq &C(\|m(t)\|_{L^1}+\|n(t)\|_{L^1})\|e^{-|x|}\|_{L^\infty}(\|m(t)\|_{L^1}+\|n(t)\|_{L^1})
\|e^{-|x|}\|_{L^\infty}(\|m(t)\|_{L^1}+\|n(t)\|_{L^1})|m(t,x)|\\
\leq &C|m(t,x)|.
\end{align*}
Following along almost the same lines as above yields
\begin{align*}
&\|-(1-\partial_x^2)^{-1}\big(\partial_x(Q_xv)+
(Q_xv_{x})\big)(t,x)m(t,x)\|_{L^\infty}\leq C|m(t,x)|,\\
&\|(1-\partial_x^2)^{-1}\big((Q_xu)-
\partial_x(Q_xu_{x})\big)(t,x)n(t,x)
-(1-\partial_x^2)^{-1}\big(\partial_x(Q_xu)-
(Q_xu_{x})\big)(t,x)n(t,x)\|_{L^\infty}\leq C|n(t,x)|.
\end{align*}
Combining the above there inequalities completes the proof of the lemma.\qed
\begin{lemm}\label{sign}
Let $m_{0},n_{0}\in H^s$ $(s>\frac{1}{2})$, and let $T>0$ be the maximal existence time of the corresponding solution $(m,n)$ to Eq.(\ref{1.2}).
 Assume that $m_{0}$ and $ n_{0}$ do not change sign. Set $Q(t,x)=\frac{1}{2}(u-u_x)(v+v_x)(t,x).$ Then there exists a constant $C=C(\|(v_0+v_{0x})m_{0}\|_{L^1},\|(u_0-u_{0x})n_{0}\|_{L^1},\|u_{0}\|_{H^1},\|v_{0}\|_{H^1})$ such that
\begin{align}
&Q_{xt}(t,x)+ \big(Q(Q_{x})_x\big)(t,x)+Q_{x}^2(t,x)\leq Ce^{Ct}(|m|+|n|)(t,x).
\end{align}
\end{lemm}
\noindent{Proof}. Applying Lemma \ref{ll1.1} to the first term on the right hand side of (\ref{a}) yields
\begin{align*}
  &(1-\partial_x^2)^{-1}\big((Q_xv)+
\partial_x(Q_xv_{x})\big)(t,x)m(t,x)\\\leq &\|(1-\partial_x^2)^{-1}\big((Q_xv)+
\partial_x(Q_xv_{x}\big)(t)\|_{L^\infty}|m(t,x)|\\
=&[\|\frac{1}{2}e^{-|x|}\ast\Big(\big(m(v+v_x)-n(u-u_x)\big)v\Big)\|_{L^\infty}\\&+
\|\frac{1}{2}\big(sign(x)e^{-|x|}\big)\ast\Big(\big(m(v+v_x)-n(u-u_x)\big)
v_x\Big)\|_{L^\infty}] |m(t,x)|\\
\leq& C (\|u-u_x\|_{L^\infty}+\|v+v_x\|_{L^\infty})(\|v\|_{L^\infty}+\|v_x\|_{L^\infty})
(\|e^{-|x|}\ast m\|_{L^\infty}+\|e^{-|x|}\ast n\|_{L^\infty})|m(t,x)|\\
=& C (\|u-u_x\|_{L^\infty}+\|v+v_x\|_{L^\infty})(\|v\|_{L^\infty}+\|v_x\|_{L^\infty})
(\|u\|_{L^\infty}+\|v\|_{L^\infty})|m(t,x)|\\\leq &Ce^{Ct}|m|(t,x),
\end{align*}
where we have used the fact that $m,$ $n$ do not change sign.
The left three terms can be treated in the same way. We have
\begin{align*}-(1-\partial_x^2)^{-1}\big(\partial_x(Q_xv)+
&(Q_xv_{x})\big)m
+(1-\partial_x^2)^{-1}\big((Q_xu)-
\partial_x(Q_xu_{x})\big)n\\
&-(1-\partial_x^2)^{-1}\big(\partial_x(Q_xu)-
(Q_xu_{x})\big)n](t,x)\leq Ce^{Ct}(|m|+|n|)(t,x).
\end{align*}
Plunging the above two inequalities into (\ref{a}) completes the proof of the lemma.\qed

Next, we present two blow-up results.
\begin{theo}\label{l1}
Let $m_{0},n_{0}\in H^s$ $(s>\frac{1}{2})$, and let $T>0$ be the maximal existence time of the corresponding solution $(m,n)$ to Eq.(\ref{1.2}). Set $Q(t,x)=\frac{1}{2}(u-u_x)(v+v_x)(t,x).$ Assume that $m_{0}$ and $ n_{0}$ do not change sign, and that there exists some $x_0\in \mathbb{R}$ such that $N(0,x_0)=|m(0,x_0)|+|n(0,x_0)|>0$ and
$Q_{x}(0,x_0)=\frac{1}{2}\big(m_0(v_0+v_{0x})-n_0(u_{0}-u_{0x})\big)(x_0)\leq a_0$, where $a_0$ is the unique negative solution to the following equation
\begin{align*}
1+ag(-\frac{a}{N(0,x_0)})+N(0,x_0)\int_0^{g(-\frac{a}{N(0,x_0)})}f(s)ds=0,
\end{align*}
with $f(x)=e^{Cx}-1, ~x\geq 0$, $g(x)=\frac{1}{C}\log(x+1), ~ x\geq 0.$\\
Then the solution $(m,n)$ blows up at a time $T_0\leq g(-\frac{Q_{x}(0,x_0)}{N(0,x_0)}).$
\end{theo}
\noindent{Proof.} In view of Lemma \ref{xinjiaL1}, we obtain that
\begin{align*}
Q_{xt}(t,x_0)+ \big(Q(Q_{x})_x\big)(t,x_0)+Q_{x}^2(t,x_0)\leq Ce^{Ct}(|m|+|n|)(t,x_0).
\end{align*}
By Lemma \ref{qq0} and Lemma \ref{ll11}, we have
\begin{align*}
&\frac{d}{dt}Q_{x}(t,q(t,x_0))+Q_{x}^2(t,q(t,x_0))\leq Ce^{Ct}(|m|+|n|)(t,q(t,x_0))=Ce^{Ct}(|m_0(x_0)|+|n_0(x_0)|)q_x^{-1}(t,x_0)\\
=&CN(0,x_0)e^{Ct}
exp(\int_0^t-Q_{x}(\tau,q(\tau,x_0))d\tau),
\end{align*}
form which it follows that
\begin{align*}
  \frac{d}{dt}\big(Q_x(t,q(t,x_0))exp(\int_0^tQ_{x}(\tau,q(\tau,x_0))d\tau)\big)\leq CN(0,x_0)e^{Ct}.
\end{align*}
Integrating from $0$ to $t$ yields
\begin{align*}
\frac{d}{dt}exp(\int_0^tQ_{x}(\tau,q(\tau,x_0))d\tau)=
Q_x(t,q(t,x_0))exp(\int_0^tQ_{x}(\tau,q(\tau,x_0))d\tau)\leq N(0,x_0)(e^{Ct}-1)+Q_{x}(0,x_0).
\end{align*}
Integrating again from $0$ to $t$ yields
\begin{align}\label{e}
(e^{\int_0^t\inf_{x\in \mathbb{R}}Q_x(\tau,x)d\tau}\leq)exp(\int_0^tQ_{x}(\tau,q(\tau,x_0))d\tau)\leq N(0,x_0)\int_0^t(e^{Cs}-1)ds+Q_{x}(0,x_0)t+1.
\end{align}
Next, we consider the following function
$$F(a,t)=1+at+N(0,x_0)\int_0^tf(s)ds, a\leq 0,$$
where $f(x)=e^{Cx}-1, ~x\geq 0$. It is easy to see that
$$\min_{t\geq 0}F(a,t)=F(a,g(-\frac{a}{N(0,x_0)}))=1+ag(-\frac{a}{N(0,x_0)})+N(0,x_0)\int_0^{g(-\frac{a}{N(0,x_0)})}f(s)ds\triangleq G(a),$$
where $g(x)=\frac{1}{C}\log(x+1), ~ x\geq 0,$ is the inverse function of $f$.
Differentiating $G(a)$ with respect to $a$, we obtain
\begin{align*}
  \frac{d}{da}G(a)=&g(-\frac{a}{N(0,x_0)})-g'(-\frac{a}{N(0,x_0)})\frac{a}{N(0,x_0)}
  +g'(-\frac{a}{N(0,x_0)})\frac{N(0,x_0)}{N(0,x_0)}\frac{a}{N(0,x_0)}\\
  =&g(-\frac{a}{N(0,x_0)})>0,~~~a< 0.
\end{align*}
Notice that
\begin{align*}
  \lim_{a\rightarrow -\infty}g(-\frac{a}{N(0,x_0)})=+\infty.
\end{align*}
Thus, we deduce that
\begin{align*}
\lim_{a\rightarrow -\infty}G(a)=-\infty,
\end{align*}
which, together with that fact that $G(0)=1$ and the continuity of $G$, yields that
there exists a unique $a_0<0$ satisfies
$G(a_0)=0.$ Therefore, $G(a)\leq 0$ if $a\leq a_0.$
Combining this with (\ref{e}), if $Q_x(0,x_0)\leq a_0,$ we may find a time $0<T_0\leq g(-\frac{Q_x(0,x_0)}{N(0,x_0)})$ such that
\begin{align*}
  e^{\int_0^t\inf_{x\in \mathbb{R}}Q_x(\tau,x)d\tau}\rightarrow 0, ~\textit{as}~t\rightarrow T_0,
\end{align*}
which, implies that
\begin{align*}
  \liminf_{t\rightarrow T}\inf_{x\in \mathbb{R}}Q_x(t,x)\rightarrow -\infty, ~\textit{as}~t\rightarrow T_0.
\end{align*}
Therefore, in view of Theorem \ref{bao222}, we conclude that the solution $(m,n)$ blows up at the time $T_0.$
\begin{theo}\label{l1l1}
Let $m_{0},n_{0}\in H^s$ $(s>\frac{1}{2})$, and let $T>0$ be the maximal existence time of the corresponding solution $(m,n)$ to Eq.(\ref{1.2}). Set $Q(t,x)=\frac{1}{2}(u-u_x)(v+v_x)(t,x).$ Assume that $m_{0} ,n_{0}\in L^1$, and that there exists some $x_0\in \mathbb{R}$ such that $N(0,x_0)=|m_0(x_0)|+|n_0(x_0)|>0$ and
$Q_{x}(0,x_0)=\frac{1}{2}\big(m_0(v_0+v_{0x})-n_0(u_{0}-u_{0x})\big)
(x_0)\leq-\big(2CN(0,x_0)\big)^{\frac{1}{2}}$.
Then there exists a constant $C=C(\|m_0\|_{L^1},\|n_0\|_{L^1})$ such that the solution $(m,n)$ blows up at a time $T_0\leq \frac{-Q_{x}(0,x_0)}{CN(0,x_0)}.$
\end{theo}
\noindent{Proof.} In view of Lemma \ref{xinjiaL1}, we obtain that
\begin{align*}
Q_{xt}(t,x)+ \big(Q(Q_{x})_x\big)(t,x)+Q_{x}^2(t,x)\leq C(|m|+|n|)(t,x).
\end{align*}
By Lemma \ref{qq0} and Lemma \ref{ll11}, we have
\begin{align*}
&\frac{d}{dt}Q_{x}(t,q(t,x))+Q_{x}^2(t,q(t,x))\leq C(|m|+|n|)(t,q(t,x))=C(|m_0(x)|+|n_0(x)|)q_x^{-1}(t,x)\\=&CN(0,x)
exp\big(\int_0^t-\frac{1}{2}\big(m(v+v_x)-n(u-u_x)\big)(\tau,q(\tau,x))d\tau\big)\\
=&CN(0,x)
exp(\int_0^t-Q_{x}(\tau,q(\tau,x))d\tau),
\end{align*}
form which it follows that
\begin{align*}
  \frac{d}{dt}\big(Q_x(t,q(t,x))exp(\int_0^tQ_{x}(\tau,q(\tau,x))d\tau)\big)\leq CN(0,x).
\end{align*}
Integrating from $0$ to $t$ yields
\begin{align*}
\frac{d}{dt}exp(\int_0^tQ_{x}(\tau,q(\tau,x))d\tau)=
Q_x(t,q(t,x))exp(\int_0^tQ_{x}(\tau,q(\tau,x))d\tau)\leq CN(0,x)t+Q_{x}(0,x).
\end{align*}
Integrating again from $0$ to $t$ yields
\begin{align*}
(e^{\int_0^t\inf_{x\in \mathbb{R}}Q_x(\tau,x)d\tau}\leq)exp(\int_0^tQ_{x}(\tau,q(\tau,x))d\tau)\leq \frac{1}{2}CN(0,x)t^2+Q_{x}(0,x)t+1.
\end{align*}
Hence, if there exists some $x_0\in \mathbb{R}$ such that $N(0,x_0)>0$ and
$Q_{x}(0,x_0)\leq-\big(2CN(0,x_0)\big)^{\frac{1}{2}},$ then we may find a time $0<T_0\leq \frac{-Q_{x}(0,x_0)}{CN(0,x_0)}$ such that
\begin{align*}
  e^{\int_0^t\inf_{x\in \mathbb{R}}Q_x(\tau,x)d\tau}\rightarrow 0, ~\textit{as}~t\rightarrow T_0,
\end{align*}
which, implies that
\begin{align*}
  \liminf_{t\rightarrow T}\inf_{x\in \mathbb{R}}Q_x(t,x)\rightarrow -\infty, ~\textit{as}~t\rightarrow T_0.
\end{align*}
Therefore, in view of Theorem \ref{bao222}, we conclude that the solution $(m,n)$ blows up at the time $T_0.$\qed
\begin{rema}
We mention that, if $v=2u$, Theorem \ref{l1} is same as Theorem 5.2 and Theorem 5.3 in \cite{Gui}, while Theorem \ref{l1l1} represents a new blow-up result for Eq.(\ref{qio}).
\end{rema}
\subsection{$N=1,H=-\frac{1}{2}(uv-u_xv_x)$}
\subsubsection{A precise blow-up scenario}
For $N=1$ and $H=-\frac{1}{2}(uv-u_xv_x)$, Eq.(\ref{s0}) is reduced to the following system:
\begin{align}\label{1.1}
\left\{
\begin{array}{l}
m_{t}+\frac{1}{2}\big((uv-u_{x}v_{x}) m\big)_x-\frac{1}{2}(uv_{x}-vu_{x})m=0,\\[1ex]
n_{t}+\frac{1}{2}\big((uv-u_{x}v_{x}) n\big)_x+\frac{1}{2}(uv_{x}-vu_{x})n=0,\\[1ex]
(m,n)|_{t=0}=(m_{0},n_{0}),
\end{array}
\right.
\end{align}
where $m=u-u_{xx}$ and $n=v-v_{xx}$.

Along the same lines as the proof of Lemma \ref{qq0}-\ref{ll11} and Theorem \ref{bao222}, we can obtain the following results.
\begin{lemm}\label{qq02}
Let $m_{10},m_{20}\in H^s~(s>\frac{1}{2})$, and let $T>0$ be the maximal existence time of the corresponding solution $M=(m_1,m_2)$ to Eq.(\ref{1.1}). Then the following system \begin{align}\label{q2}
\left\{
\begin{array}{l}
q_{t}(t,x)=\frac{1}{2}(uv-u_{x}v_{x})(t,q),~~~t\in[0,T),\\[1ex]
q(0,x)=x, ~~~x\in \mathbb{R}.
\end{array}
\right.
\end{align} has a unique solution $q\in C^1([0,T]\times \mathbb{R};\mathbb{R}).$ Moreover, the mapping $q(t,\cdot)$ $(t\in[0,T))$ is an increasing diffeomorphism of $\mathbb{R},$ with
\begin{align}\label{qx2}
q_x(t,x)=exp\big(\int_0^t\frac{1}{2}(mv_x+nu_x)(\tau,q(\tau,x))d\tau\big).
\end{align}
\end{lemm}
\begin{lemm}\label{ll112}
  Let $m_{10},m_{20}\in H^s~(s>\frac{1}{2})$, and let $T>0$ be the maximal existence time of the corresponding solution $M=(m_1,m_2)$ to Eq.(\ref{1.2}). Then, we have for all $t\in[0,T),$
  \begin{align}
    &m(t,q(t,x))q_x(t,x)=m_{0}(x)exp\big(\frac{1}{2}\int_0^t(uv_{x}-vu_{x})(\tau,q(\tau,x))d\tau\big),\label{m112}\\
    &n(t,q(t,x))q_x(t,x)=n_{0}(x)exp\big(-\frac{1}{2}\int_0^t(uv_{x}-vu_{x})(\tau,q(\tau,x))d\tau\big).\label{m222}
  \end{align}
\end{lemm}
\begin{theo}\label{bao2222}
   Let $m_{0},n_{0}\in H^s$ $(s>\frac{1}{2})$, and let $T>0$ be the maximal existence time of the corresponding solution $(m,n)$ to Eq.(\ref{1.1}). Then the solution $(m,n)$ blows up in finite time if and only if
   \begin{align*}
     \liminf_{t\rightarrow T}\inf_{x\in \mathbb{R}}\big((mv_x+nu_x)\big)(t,x)=-\infty~~~\textit{or}~~~\limsup_{t\rightarrow T}\|(uv_x-vu_x)(t,\cdot)\|_{L^\infty}=+\infty.
   \end{align*}
\end{theo}
\subsubsection{Blow-up phenomena}
Now we derive four useful conservation laws for Eq.(\ref{1.1}).
\begin{lemm}\label{L02}
Let $m_{0},n_{0}\in H^s$ with $s>\frac{1}{2},$ and let $T>0$ be the maximal existence time of the corresponding solution $(m,n)$ to Eq.(\ref{1.1}). Then we have that for all $t\in[0,T),$
\begin{align*}
  \int_{\mathbb{R}}(mv_x)(t,x)dx&=\int_{\mathbb{R}}m_{0}v_{0x}dx,
 \int_{\mathbb{R}}(nu_x)(t,x)dx=\int_{\mathbb{R}}n_{0}u_{0x}dx,\\
 \int_{\mathbb{R}}(mv)(t,x)dx&=\int_{\mathbb{R}}m_{0}v_{0}dx,
 \int_{\mathbb{R}}(nu)(t,x)dx=\int_{\mathbb{R}}n_{0}u_{0}dx.
\end{align*}
\end{lemm}
\noindent{Proof}. By Eq.(\ref{1.2}), we have
\begin{align*}
  &\frac{d}{dt}\int_{R}(mv_x)(t,x)dx=\frac{d}{dt}\int_{R}(-nu_x)(t,x)dx\\
  =&\int_{\mathbb{R}}\big(v_xm_{t}-u_xn_{t}\big)(t,x)dx\\
  =&\frac{1}{2}\int_{\mathbb{R}}[(uv-u_xv_x)\big(mv_{xx}-nu_{xx}\big)(t,x)
  +(uv_{x}-vu_{x})\big(v_xm+u_xn\big)]dx\\
  =&\frac{1}{2}\int_{\mathbb{R}}\partial_x\big((uv-u_xv_x)(uv_{x}-vu_{x})\big)(t,x)dx
  =0,
\end{align*}
and \begin{align*}
  &\frac{d}{dt}\int_{\mathbb{R}}(mv)(t,x)dx=\frac{d}{dt}\int_{\mathbb{R}}(nu)(t,x)dx
  =\int_{\mathbb{R}}(m_{t}v+n_{t}u)(t,x)dx\\
  =&\frac{1}{2}\int_{\mathbb{R}}\big((uv-u_{x}v_{x})(mv_{x}+nu_{x})+(uv_{x}-vu_{x})
  (mv-nu)\big)(t,x)dx\\
  =&\frac{1}{2}\int_{\mathbb{R}}\big((uv-u_{x}v_{x})\partial_x(uv-u_{x}v_{x})-
  (uv_{x}-vu_{x})\partial_x(uv_{x}-vu_{x})\big)(t,x)dx=0.
\end{align*}
This completes the proof of the lemma.\qed
\begin{lemm}\label{l1.2}
Let $m_{0},n_{0}\in H^s$ $(s>\frac{1}{2})$, and let $T>0$ be the maximal existence time of the corresponding solution $(m,n)$ to Eq.(\ref{1.1}). Assume that $m_{0}$ and $n_{0}$ do not change sign. Then there exists a constant $C=C(\|v_{0x}m_{0}\|_{L^1},\|v_{0}m_{0}\|_{L^1},\|u_{0x}n_{0}\|_{L^1},\|u_{0}n_{0}\|_{L^1},\|u_{0}\|_{H^1},\|v_{0}\|_{H^1})$ such that
\begin{align}
|u_{x}(t,x)|\leq |u_{}(t,x)|,~~~|v_{x}(t,x)|\leq |v_{}(t,x)|,\\
  \|u(t)\|_{H^1}+\|v(t)\|_{H^1}\leq Ce^{Ct},~\forall t\in[0,T).\label{uu12}
  \end{align}
\end{lemm}
\noindent{Proof.} Without loss of generality, we assume that $m_{0}\geq 0,$ $n_{0}\geq 0$. Repeating the arguments that were used in Lemma \ref{ll1.1}, we get that the inequalities (\ref{m11})-(\ref{u2}) still hold true here. Next, according to Lemma \ref{L02} with $m,u+u_x,n,v-v_x\geq 0$, we obtain
\begin{align}
  \|(mv_{x})(t)\|_{L^1}&\leq\|\big(m(v-v_{x})\big)(t)\|_{L^1}+\|(mv)(t)\|_{L^1}\\\nonumber
  &=\int_{\mathbb{R}}\big(m(v-v_{x})\big)(t,x)dx
  +\int_{\mathbb{R}}(mv)(t,x)dx\\\nonumber
  &=2\int_{\mathbb{R}}(mv)(t,x)dx-\int_{\mathbb{R}}mv_x(t,x)dx\\\nonumber
  &\leq 2\|(m_0v_0)\|_{L^1}+\|(m_0v_{0x})\|_{L^1}.
\end{align}
Finally, form Eq.(\ref{1.1}), we have
\begin{align*}
  \frac{1}{2}\frac{d}{dt}(\|u(t)\|_{H^1}^2+\|v(t)\|_{H^1}^2)
  =&\int_{\mathbb{R}}(m_{t}u+n_{t}v)(t,x)dx\\
=&\frac{1}{2}\int_{\mathbb{R}}\big((uv-u_{x}v_{x})mu_{x}+(uv_{x}-vu_{x})mu\\&+
  (uv-u_{x}v_{x})nv_{x}-(uv_{x}-vu_{x})nv\big)(t,x)dx\\
  =&\frac{1}{2}\int_{\mathbb{R}}\big((u^1-u_{x}^2)v_{x}m+(v^2-v_{x}^2)u_{x}n\big)(t,x)dx\\
  \leq &\frac{1}{2}(\|(u^2-u_{x}^2)(t)\|_{L^\infty}\|(mv_{x})(t)\|_{L^1}+
  \|(v^2-v_{x}^2)(t)\|_{L^\infty}\|(nu_{x})(t)\|_{L^1}\\
  \leq &C(\|u(t)\|_{L^\infty}^2+\|v(t)\|_{L^\infty}^2)
  \leq C(\|u(t)\|_{H^1}^2+\|v(t)\|_{H^1}^2).
\end{align*}
Then the Gronwall lemma yields the desired inequality (\ref{uu12}). This completes the proof of the lemma.\qed
\begin{theo}\label{l22}
Let $m_{0},n_{0}\in H^s$ $(s>\frac{1}{2})$, and let $T>0$ be the maximal existence time of the corresponding solution $(m,n)$ to Eq.(\ref{1.1}). Set $Q(t,x)=\frac{1}{2}(uv-u_xv_x)(t,x)$. Assume that $m_{0} ,n_{0}$ do not change sign, and that there exists some $x_0\in \mathbb{R}$ such that $N(0,x_0)=|m(0,x_0)|+|n(0,x_0)|>0$ and
$Q_x(0,x_0)=\frac{1}{2}\big(m_0v_{0x}+n_0u_{0x})\big)(x_0)\leq a_0$, where $a_0$ is the unique negative solution to the following equation
\begin{align*}
1+ag(-\frac{a}{N(0,x_0)})+N(0,x_0)\int_0^{g(-\frac{a}{N(0,x_0)})}f(s)ds=0,
\end{align*}
with $f(x)=exp(e^{Cx}-1)-1, ~x\geq 0$, $g(x)=\frac{1}{C}\log\big(\log(x+1)+1\big), ~ x\geq 0.$\\
Then the solution $(m,n)$ blows up at a time $T_0\leq g(-\frac{Q_x(0,x_0)}{N(0,x_0)}).$
\end{theo}
\noindent{Proof.}  It follows from Eq.(\ref{1.1}) that
\begin{align}\label{b}
&Q_{xt}+Q(Q_{x})_x+Q_{x}^2\\\nonumber=
&
-(1-\partial_x^2)^{-1}\Big(\partial_x(Q_xu)+
(Q_xu_{x})-\partial_x\big(\frac{1}{2}(uv_x-uv_x)m\big)\Big)n\\\nonumber
&-(1-\partial_x^2)^{-1}\Big(\partial_x(Q_xv)+
(Q_xv_{x})+\partial_x\big(\frac{1}{2}(uv_x-uv_x)n\big)\Big)m\\\nonumber
&+\frac{1}{2}(uv_x-vu_x)(mv_{x}-nu_{x}).
\end{align}
Using Lemma \ref{l1.2}, and following along the same lines as the proof of Lemma \ref{sign}, we obtain that
\begin{align*}
Q_{xt}(t,x_0)+ \big(Q(Q_{x})_x\big)(t,x_0)+Q_{x}^2(t,x_0)\leq Ce^{Ct}(|m|+|n|)(t,x_0).
\end{align*}
By Lemma \ref{ll112}, we get \begin{align*}
&\frac{d}{dt}Q_{x}(t,q(t,x_0))+Q_{x}^2(t,q(t,x_0))\leq Ce^{Ct}(|m|+|n|)(t,q(t,x_0))\\\leq&Ce^{Ct}N(0,x_0)
exp\big(\int_0^t-\frac{1}{2}\big(m(v+v_x)-n(u-u_x)\big)(\tau,q(\tau,x_0))d\tau\big)
exp\big(\frac{1}{2}\int_0^t\|(uv_{x}-vu_{x})(\tau)\|_{L^\infty}d\tau\big)\\=
&Ce^{Ct}N(0,x_0)
exp(\int_0^t-Q_{x}(\tau,q(\tau,x_0))d\tau)
exp\big(\frac{1}{2}\int_0^t\|(uv_{x}-vu_{x})(\tau)\|_{L^\infty}d\tau\big).
\end{align*}
Again using Lemma \ref{l1.2}, we have
\begin{align*}
exp\big(\frac{1}{2}\int_0^t\|(uv_{x}-vu_{x})(\tau)\|_{L^\infty}d\tau\big)
\leq exp(C \int_0^te^{C\tau} d\tau)=exp(e^{Ct}-1),
\end{align*}
from which it follows that
\begin{align*}
   \frac{d}{dt}\big(Q_x(t,q(t,x_0))exp(\int_0^tQ_{x}(\tau,q(\tau,x_0))d\tau)\big)\leq Ce^{Ct}exp(e^{Ct}-1)N(0,x_0).
\end{align*}
Integrating from $0$ to $t$ yields
\begin{align*}
  \frac{d}{dt}e^{\int_0^tQ_x(\tau,q(\tau,x_0))d\tau}=&e^{\int_0^tQ_x(\tau,q(\tau,x_0))d\tau} Q_x(t,q(t,x_0))
  \leq Q_x(0,x_0)+N(0,x_0)\int_0^texp(e^{C\tau}-1)Ce^{C\tau}d\tau\\
  =&Q_x(0,x_0)+N(0,x_0)(exp(e^{Ct}-1)-1).
\end{align*}
Integrating again from $0$ to $t$ yields
\begin{align}\label{e2}
 ( e^{\int_0^t\inf_{x\in \mathbb{R}}Q_x(\tau,x)d\tau}\leq) e^{\int_0^tQ_x(\tau,q(\tau,x_0))d\tau}\leq1+
 Q_x(0,x_0)t+N(0,x_0)\int_0^t(exp(e^{C\tau}-1)-1)d\tau.
\end{align}
Next, following along almost the same lines as in the proof of Lemma \ref{l1} with
$f(x)=exp(e^{Cx}-1)-1, ~x\geq 0$ and $g(x)=\frac{1}{C}\log\big(\log(x+1)+1\big), ~ x\geq 0,$ completes the proof of the theorem.\qed
\begin{rema}
We mention that Theorem \ref{l22} is an improvement of Theorem 4.3 in \cite{yan}. Firstly, in \cite{yan} the authors assumed that  $\|u\|_{L^\infty},\|v\|_{L^\infty}\leq Ce^{Ct}$, while in our paper, $\|u\|_{L^\infty},\|v\|_{L^\infty}\leq Ce^{Ct}$ is ensured by Lemma \ref{l1.2}. Secondly, in \cite{yan} $x_0$ is required to satisfy an additional restriction: $Q_x(0,x_0)=\inf_{x\in\mathbb{R}}Q_x(0,x)$. Finally, $a_0$ in our result is more explicit and accurate than that in \cite{yan}.
\end{rema}

\vspace*{2em} \noindent\textbf{Acknowledgements}. This work was
partially supported by NNSFC (No.11271382), RFDP (No.
20120171110014), and the key project of Sun Yat-sen University.

\bibliography{reference}

\end{document}